\RequirePackage{xr-hyper}
\documentclass{fun_article}

\title{Attention shapes collective decisions for the Antarctic global commons}

\definecolor{orcidlogocol}{HTML}{A6CE39}

\author[*, 1, 2, 3]{Casper van Elteren\orcid{0000-0001-9862-8936}}
\author[5]{Fiona Lippert\orcid{0000-0003-4174-2230}}
\author[1]{Zach Carter\orcid{0000-0002-0709-4412}}
\author[1]{Maria Kleshnina\orcid{0000-0002-5518-8317}}
\author[1]{Michael Bode\orcid{0000-0002-5886-4421}}
\author[3, 4]{Vítor V. Vasconcelos\orcid{0000-0002-4621-5272}}

\affil[*]{Corresponding author: casper.vanelteren@qut.edu.au}
\affil[1]{Securing Antarctica's Environmental Future, School of Mathematical Sciences, Queensland University of Technology, Brisbane, Australia}
\affil[2]{Institute for Advanced Study, University of Amsterdam, 1012 GC Amsterdam, Netherlands}
\affil[3]{Computational Science Lab, Informatics Institute, University of Amsterdam, 1090 GH Amsterdam, The Netherlands}
\affil[4]{Polder Center, Institute for Advanced Study, University of Amsterdam, 1012 GC Amsterdam, The Netherlands}
\affil[5]{Space Research Organization Netherlands, 2333 CA Leiden, The Netherlands}
\newcommand{\significancetext}{%
    Planetary problems such as climate change fall to states that answer to no common authority and must agree among themselves. Antarctica, whose ice and ocean shape global climate and sea level, has been governed this way for six decades. Its agreements record what all parties accept, not where attention went beforehand. Borrowing tools economists use to study export diversification, we map this attention, recorded in papers submitted to Treaty meetings, as a network of concerns. Participants concentrate on different parts of a shared agenda and expand into closely linked concerns. Attention helps predict the subjects of later agreements, beyond recent agreements alone. When no one is in charge, agreements show what has been settled. Attention anticipates what they address next.
}

\sidebar{%
    \begin{significancebox}
    \significancetext
    \end{significancebox}
}

\makeatletter
\@ifundefined{if@kernel@version@MMXXV}{\newif\if@kernel@version@MMXXV}{}
\let\@setmarks\@empty
\makeatother



\makeatletter
\begin{document}
\makeatother
\sloppy

\begin{abstract}
Consensus-based regimes govern critical global commons, yet this need for unanimity obscures how agendas develop before formal policies are enacted and who cares about what aspects of the commons. We examine 6,573 papers submitted by 66 participants to Antarctic Treaty Consultative Meetings from 1961 to 2025. From these papers, we reconstruct a network of concerns that maps the structure of collective attention within the treaty system.
Participants concentrate their concerns on different parts of this network and tend to develop new specializations in concerns closely linked to their earlier work. Following participants’ contributions reveals a historically structured process of agenda development that results in new agreements. This concern structure from papers is associated with subjects in effectively adopted agreements, predicting agreement subjects beyond what previous agreements alone can capture. Participants follow different paths through the same shared agenda, and their earlier specializations help anticipate the concerns in which they develop new specializations. As pressures from tourism, living resource extraction, and environmental change intensify, these findings show how participants' histories shape and constrain the capacity to support pressing emerging concerns.
\end{abstract}


\section{Introduction}
In August 2026, average sea-surface temperature across the extra-polar oceans reached its highest daily value in the observational record \cite{Copernicus2026OceanRecord}. The measurement stops at $60^\circ$S latitude, but the processes moving heat through the ocean do not. As United Nations Secretary-General Ant\'onio Guterres put it after visiting the continent, ``What happens in Antarctica doesn't stay in Antarctica'' \cite{Guterres2023Antarctica}. Southern Ocean circulation strongly influences global heat storage, anthropogenic carbon uptake, overturning circulation, and sea level \cite{IPCC2021OceanCryosphere,Holland2020}. Antarctica, therefore, sits near the center of a climate system whose consequences extend far beyond the continent. Yet, it has no permanent population and is governed not by a single sovereign authority but through treaties among states \cite{AntarcticTreaty1959,Joyner1998}. Antarctica, thus, poses a collective governance problem with planetary consequences.

How Antarctic governance responds to global change depends in part on which concerns receive collective attention. Since institutional attention is limited, the way information is processed helps determine which problems receive priority \cite{JonesBaumgartner2005}. In international governance, the work that establishes these priorities is distributed across states and organizations, whose relationships structure collective activity \cite{Kim2020}. Which concerns receive attention, therefore, depends on how these actors bring them into discussion, and it is visible, though only partly, in the agreements eventually reached \cite{GutnerThompson2010}.

The Antarctic Treaty System provides a long documentary record of this process. The Treaty, signed in 1959 and in force since 1961, established peaceful use and scientific cooperation as foundations of governance on the continent \cite{AntarcticTreaty1959,Joyner1998}. At Antarctic Treaty Consultative Meetings (ATCMs), Treaty Parties, observers, the Secretariat, and invited experts, collectively termed ``participants'', submit papers that bring concerns and proposed responses into discussion \cite[pp.~3, 7--8]{ATSDelegatesManual2025}. Here, a concern is one of the 45 categories under which the Secretariat's archive classifies submitted papers, such as impact assessment, tourism, or inspections. We measure a participant’s \emph{attention} to a concern by the papers it submits on that concern; the attention a concern receives is the total across participants. However, papers are not agreements. Only Consultative Parties adopt agreements, which take three forms (Measures, Resolutions, and Decisions), and they do so by consensus \cite{ATCMRules2015,HughesEtAl2024ConsultativeStatus}. Comparing submitted papers with adopted agreements makes it possible to trace both the concerns participants raise and those on which collective agreement is reached \cite{Gardiner2025}. We call the concerns receiving attention through submitted papers, together with how that attention is distributed among them, the \emph{shared agenda}. It is shared in that all participants contribute to it, not in that they attend to the same concerns. \emph{Agenda development} is the changes in how attention is distributed among participants and concerns. We do not use \emph{shared agenda} for the formal list of items considered at each meeting, which we do not analyze, and it is also distinct from the specific subjects of adopted agreements, which we analyze separately.

We use \emph{specialization} to describe a participant's relative emphasis on a concern (a participant may submit papers on a concern without meeting the specialization threshold), and \emph{portfolio} for the set of concerns in which it specializes. When the same participant specializes in two concerns, for example, environmental impact assessment and visitor guidelines, we say it \emph{co-specializes} in them. Two concerns are then linked more strongly the larger the share of each concern's specializing participants that also specialize in the other. \emph{Co-specialization}, therefore, connects concerns within the shared agenda, even when participants submit papers separately. Paper counts for each concern describe how much attention it receives, but they cannot distinguish whether the same participants specialize in two concerns or separate groups work on each. Linking concerns through co-specialization seeks to highlight underlying interests and, therefore, complements network approaches that connect participants, institutions, or agreements \cite{Kim2013,Kim2020} by showing how work on different concerns overlaps.

We ask whether the structure revealed by co-specialization helps explain how the shared agenda develops and how it relates to effectively adopted agreements. Knowledge and experience by participants gained through earlier work on one concern can be useful in addressing another, but their usefulness depends on how the concerns are related \cite{Hidalgo2007, Guevara2016}. For example, experience assessing how human activities disturb wildlife may help a participant develop proposals for visitor guidelines. If co-specialization captures these relationships, participants should develop new specializations in concerns closely linked to those already in their portfolios. Since submitted papers bring information and proposals into collective discussion, attention to a concern may also help anticipate the subjects of adopted agreements, and attention to connected concerns may add further information.

Here, we adapt the shared-specialization approach of Hidalgo and colleagues \cite{Hidalgo2007} to construct a \emph{network of concerns} from papers submitted to ATCMs between 1961 and 2025, linking concerns more strongly the greater their co-specialization. We describe where participants specialize within this network and test whether their new specializations, just one type of agenda development, arise in concerns closely linked (in this network of concerns) to their earlier portfolios. We, then, turn from individual portfolios to actually adopted agreements. Because the official register of adopted agreements classifies agreements into fewer and broader subjects than the concerns used for papers, we group concerns into these subjects and build a coarser network in the same way. We test whether attention to a subject helps anticipate its share of adopted agreements beyond agreements adopted at recent meetings, and whether attention to subjects connected to it in the coarser network adds further information. Both tests rebuild their network from earlier records only.

We find that participants specialize in different parts of a connected Antarctic agenda and tend to develop new specializations in concerns closely linked to their earlier portfolios. Attention to concerns also improves retrospective predictions of the subject shares of potentially binding Measures and nonbinding Resolutions relative to a model based on recent agreements.
\begin{figure*}[!t]
\centering \includegraphics[width=\textwidth]{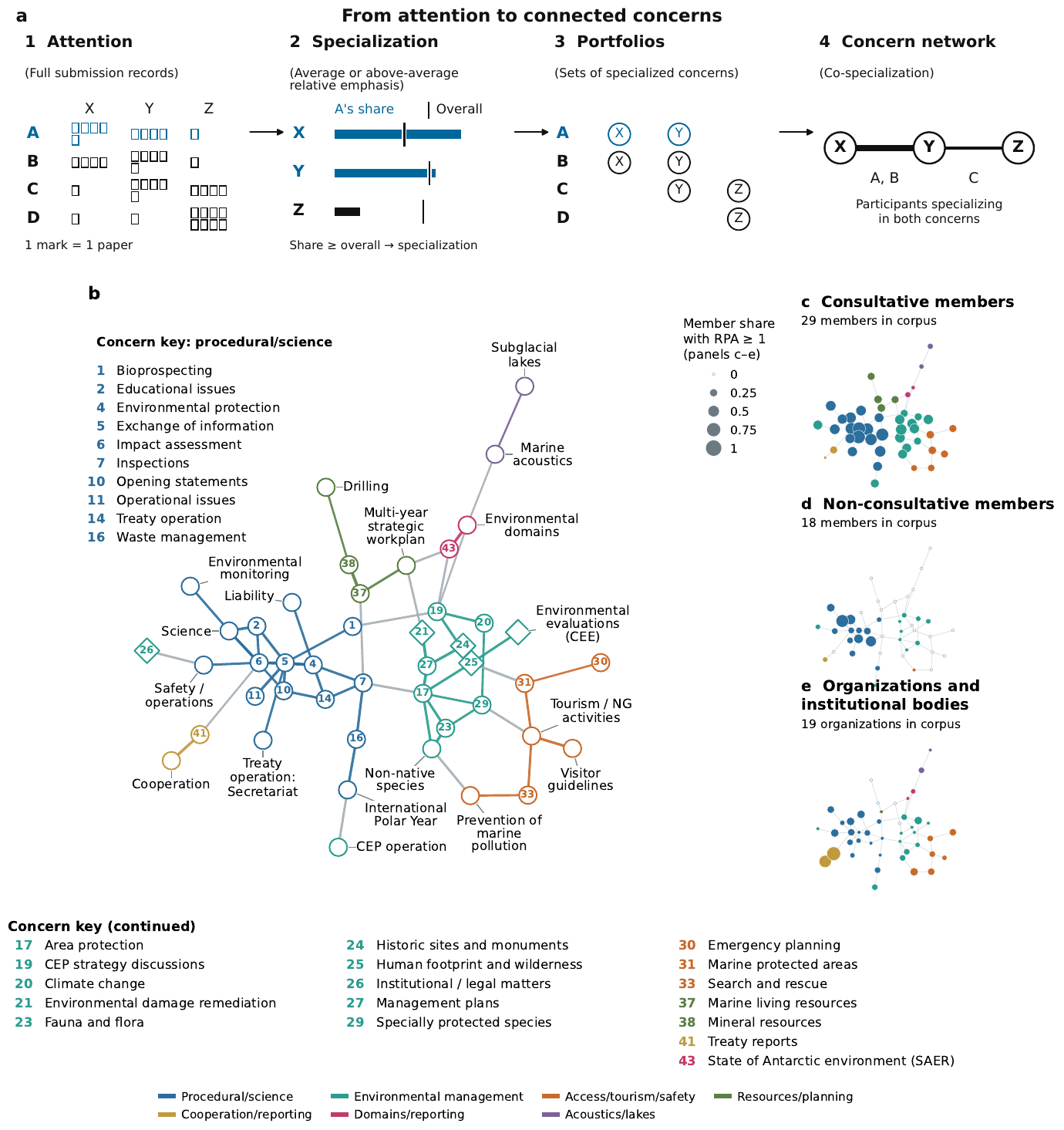}
\caption{\textbf{Co-specialization connects concerns across the shared governance agenda.} (a) Illustrative construction from paper submission records to specialization portfolios and links between concerns. (b) The network of concerns represents the emergent organization of co-specialization activity within participants' portfolios, across participants. Colors identify seven regions of related work; thicker links indicate greater overlap among specializing participants. Links form a maximum spanning tree supplemented by bootstrap-supported connections (see Methods), and diamonds mark uncertain region assignments. (c--e) The same network shows the distribution of specialization among Consultative Parties, Non-Consultative Parties, and organizations and institutional bodies. Circle area gives the proportion of each group's represented participants specializing in a concern, on a common scale. Hollow circles indicate concerns in which no participant in the group specializes.}
\label{fig:the_space_of_concerns}
\end{figure*}

\section{Results}
The Antarctic Treaty provides a framework for peaceful use, scientific cooperation, and information exchange, and includes provisions for conserving living resources \cite{AntarcticTreaty1959}. Environmental protection subsequently developed through further treaties, including the Environmental Protocol \cite{MadridProtocol1991}. We examine how participants organize their attention around the concerns addressed within this framework.

Submitted papers provide a record of this attention (\cref{fig:the_space_of_concerns}a). Attention refers to the full submission record and specialization to a participant's relative emphasis on particular concerns. To measure specialization, we compare a concern's share of a participant's papers with its share across all participants. A ratio of at least one (\textbf{R}evealed \textbf{P}olicy \textbf{A}dvantage, RPA $\ge 1$), the specialization threshold, identifies average or above-average relative emphasis.

The network connects concerns through co-specialization, showing which pairs of concerns the same participants co-specialize in (\cref{fig:the_space_of_concerns}b). Stronger connections indicate greater overlap among the participants specializing in each concern (see \cref{SI-sec:map-construction} for the definition of link weight).
Grouping concerns with relatively strong connections gives the seven colored regions shown in the figure (\cref{SI-sec:cpm-reading}). These provide a coarse description of the network, which is more clustered than expected from the numbers of specializations per participant and per concern (\cref{SI-sec:fixed-margin-null}).

The three largest regions give the network much of its shape. The procedural and scientific region brings together science, information exchange, Treaty operations, and impact assessment, alongside environmental monitoring and waste management. The environmental-management region includes area protection, management plans, species protection, and climate change. A third region brings together tourism, visitor guidelines, and search and rescue, with connections to environmental management. Tourism brings commercial activity into an agenda centered on scientific cooperation and conservation. These concentrations make visible the combinations of documentary work through which participants take up the institution's broad purposes (\cref{SI-tab:concern-case-profiles}). Smaller regions contain more focused combinations of work on resources and planning, cooperation and reporting, environmental domains and reporting, and specialized scientific concerns.

In the displayed network, the connection between inspections and area protection links the procedural and scientific region to environmental management. Inspections also connect this region to resource concerns (\cref{SI-fig:concern-network-diagnostics}b,c; \cref{SI-sec:bridge-stability}). Established under Article VII of the Antarctic Treaty, inspections provide for mutual oversight of activities in Antarctica. Their connection to area protection brings Treaty oversight and environmental-management work together within participants' portfolios.

The network of concerns offers a bottom-up perspective on governance through participants' contributions. It reveals how their combined work is organized and provides a common structure for examining \emph{where} individual participants specialize.

\paragraph{One shared agenda embeds different portfolios.}
Specialization is distributed differently across participant groups (\cref{SI-sec:group-positioning-check}). Comparing the proportion of participants specializing in each concern shows that Consultative Parties specialize in concerns throughout the network, with specialization most widespread in the procedural and scientific region and environmental management (\cref{fig:the_space_of_concerns}c). Among the Non-Consultative Parties represented in the corpus, specialization is concentrated more narrowly in science and Treaty procedures, together with environmental monitoring and protection. None specialize in tourism or resource concerns (\cref{fig:the_space_of_concerns}d).

Organizations and institutional bodies show a different pattern, with specialization most widespread in cooperation and Treaty reporting. Their specializations also extend into tourism and safety and environmental concerns such as climate change and marine pollution (\cref{fig:the_space_of_concerns}e). The panels thus show different concentrations of work within the same connected agenda.
\begin{figure*}[!t]
 \centering
 \includegraphics[width=0.95\textwidth]{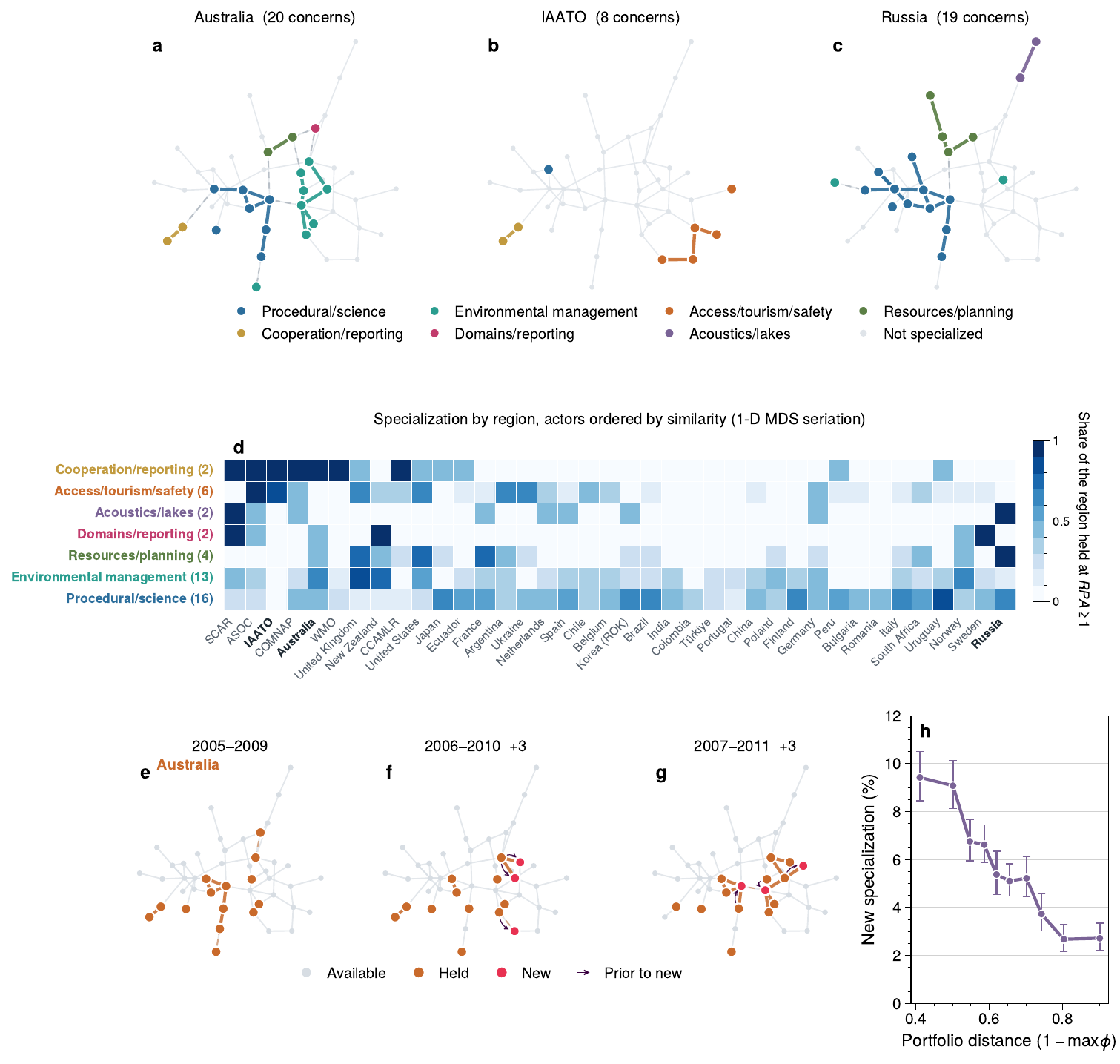}
\caption{\textbf{Participants have distinct portfolios and develop new specializations in closely linked concerns.} (a--c) Example portfolios: Australia, the International Association of Antarctica Tour Operators, and Russia. (d) Proportion of concerns specialized in within each region per participant. Participants are not active across the entire portfolio, but show differentiation within the groups indicated in \cref{fig:the_space_of_concerns} c--e. (e--g) Australia's rolling five-meeting portfolios. Pink circles mark new specializations; arrows point to them from the most closely linked concern in the earlier portfolio. (h) Percentage of previously unspecialized concerns becoming specializations, grouped by distance to earlier portfolios. Distance is one minus the strongest link to any concern in the earlier portfolio. Bars show Wilson 95\% intervals. Definitions, sample sizes, and adjusted estimates appear in \cref{SI-sec:hazard-model}.}
 \label{fig:local-specialization}
\end{figure*}

\paragraph{New specializations arise in closely linked concerns.}
Individual portfolios illustrate these different concentrations of work. Australia specializes in procedural, environmental, resource, cooperation, and reporting work, while Russia's portfolio centers on the procedural core and extends into resources and environmental concerns. The International Association of Antarctica Tour Operators specializes more narrowly in access, tourism, and safety (\cref{fig:local-specialization}a--d).

Across the archive, participants tend to develop new specializations in concerns closely linked to their earlier portfolios. Australia illustrates this pattern across rolling five-meeting windows (\cref{fig:local-specialization}e--g). Earlier specializations therefore help predict where participants develop new specializations. The rate of new specialization falls from $9.4\%$ in the lowest distance decile to $2.7\%$ in the highest (\cref{fig:local-specialization}h). After accounting for how many participants already specialize in the concern, each $0.1$ increase in distance corresponds to $18.3\%$ lower odds of specialization (OR $0.817$, 95\% model-based interval $[0.783,0.853]$; \cref{SI-sec:hazard-model}). This result persists across different time periods, network construction methods, and definitions of new specialization (\cref{SI-sec:substantive-sensitivity,SI-sec:paired-null-robustness,SI-tab:locality_estimands}). A new specialization added to the portfolio marks a change in relative emphasis and need not coincide with a participant's first paper on that concern.

\begin{figure*}[!t]
 \centering
 \includegraphics[width=\textwidth]{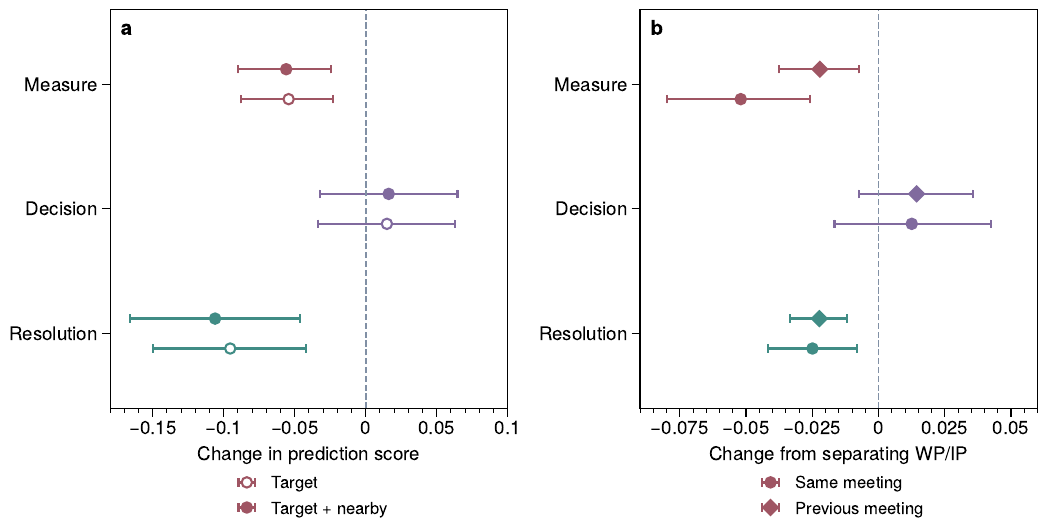}
\caption{\textbf{Attention to subjects from submitted papers improves predictions of the subjects addressed by Measures and Resolutions.} (a) Change in prediction score when submitted papers are added to a reference using agreements from the preceding five meetings. Open circles use papers addressing the predicted subject, whereas filled circles also include attention to connected subjects. (b) Further score change when Working Papers and Information Papers are counted separately rather than together. Circles use target- and preceding-meeting papers; diamonds use preceding-meeting papers alone. Predictions are of subject shares given the observed number of agreements of each form (Measures, Resolutions, or Decisions). Negative values indicate improvement. Whiskers show paired meeting-bootstrap 95\% intervals over 19 meetings from 2006 to 2025, without adjustment for retrospective model selection. Specifications appear in \cref{SI-sec:prospective-output-forecast,SI-sec:document-type-prediction}.}
 \label{fig:selective-translation}
\end{figure*}

\subsection{Attention helps anticipate the subjects of Measures and Resolutions}
The preceding analyses show where participants specialize and how their portfolios develop over time. We now ask whether the attention recorded in submitted papers also helps better anticipate the subjects of adopted agreements beyond what recent agreements alone anticipate.

The submitted papers have different roles in this process. Working Papers address matters requiring discussion and action, while Information Papers provide supporting material relevant to the meeting \cite{ATCMRules2015,ATSDelegatesManual2025}. Counting these separately lets us test whether their different roles provide information that is lost when all submitted papers are treated alike.

Adopted agreements also take different forms. Measures become legally binding after the required approvals, Resolutions express nonbinding recommendations, and Decisions address internal organizational matters \cite{ATCMDecision1995Instruments}. We analyze each form separately. To compare submissions with agreements, we group the 45 paper concerns into the 15 subjects used in the official instrument register, such as Area protection and management (\cref{SI-sec:outcome-positioning}). Given the observed total number of agreements of each form adopted at a meeting, we predict how those agreements are distributed across subjects. The target is, therefore, the subject shares of adopted agreements, rather than whether a particular proposal is adopted.

Our reference model uses agreements of the same form from the preceding five meetings. We then test whether adding Working Paper and Information Paper counts, together with the number of contributing participants, improves these predictions. A further comparison adds attention to connected subjects, using the coarser network. This distinguishes information in papers addressing a subject directly from information in attention to connected subjects.

For each meeting, models are fitted using earlier meetings. The primary comparison includes papers from the meeting being evaluated and its preceding meeting; a separate comparison uses only preceding-meeting papers. We assess how closely predicted subject shares match the observed shares using a negative-log prediction score, for which lower values indicate better predictions (\cref{SI-sec:prospective-output-forecast}).

Attention to subjects from submitted papers improves predictions of the subject shares of Measures and Resolutions beyond what recent agreements alone reveal, while Decisions show no clear improvement (\cref{fig:selective-translation}a). Separating Working Papers and Information Papers improves predictions for Measures (score change $-0.052$) and Resolutions ($-0.025$), with no clear improvement for Decisions ($+0.013$). Both the Measure and Resolution gains persist when only papers from the preceding meeting are used (\cref{fig:selective-translation}b). Using preceding-meeting papers, Working Paper attention receives a positive coefficient for Measures and Information Paper attention a negative one; both receive positive coefficients for Resolutions (\cref{SI-tab:document-type-coefficients}). Resolution and Measure gains also persist when submitted papers are restricted to Consultative Parties (\cref{SI-sec:consultative-party-sensitivity}).

Once papers addressing a subject directly are included, the additional contribution of attention to connected subjects is minimal. The gains also depend on how agreement history is represented and are less clear against a reference based on all earlier agreements (\cref{SI-sec:document-type-prediction,SI-tab:document-type-references}). Model choices were informed by the evaluation meetings, so these comparisons remain retrospective.

The concentration of Measures in area-protection and management work helps explain why earlier agreements are informative about later agreement subjects (\cref{SI-fig:measure-composition}). Most Measures also cite earlier Measures (\cref{SI-fig:measure-formal-citations}), providing a record of formal continuity. The prediction gain from attention to subjects in submitted papers shows that this continuity can coexist with additional information in participants' contributions.

\section{Discussion}
Participants maintain a shared Antarctic agenda through different but overlapping specializations. New specializations tend to develop in concerns closely linked to earlier portfolios, and attention to subjects from submitted papers adds information about the subjects of adopted Measures and Resolutions beyond recent agreements alone. Together, these findings connect the organization and development of participants' work to the subjects of adopted agreements.

Existing accounts explain how attention shapes the selection of problems for decision \cite{JonesBaumgartner2005}. Network approaches reveal how connections among agreements and institutions structure global governance \cite{Kim2013,Kim2020}. We extend this relational perspective to the shared agenda within a governance forum. Paper totals identify which concerns receive attention \cite{Gardiner2025}, but understanding how that attention is organized also requires knowing whether the same participants specialize in different concerns. Linking concerns through co-specialization of each participant reveals how concerns overlap within participants' portfolios, showing how a diverse shared agenda is sustained through participants' separate contributions (\cref{fig:the_space_of_concerns}b; \cref{fig:local-specialization}a,c).

Participants tend to develop new specializations in concerns closely linked to those already in their portfolios, making earlier work informative about where new specialization develops next. This extends relatedness findings in economic activity and scientific research to documentary work within multilateral governance \cite{Neffke2011,Hidalgo2018,Guevara2016}. A concern's presence on the shared agenda does not make it equally likely to become a specialization for every participant. Understanding how the shared agenda develops, therefore, requires considering the histories of the participants contributing to it.

In a consensus-based forum, adopted agreements record the subjects on which the collective agreement was reached, while submitted papers retain a broader record of participants' work. Gardiner and colleagues found that Antarctic discussion remained broad while the subjects of potentially binding agreements narrowed \cite{Gardiner2025}. This distinction is visible in our record, where Measures concentrate on recurring area-protection and management work. Attention in submitted papers nevertheless adds information about agreement subjects beyond recent agreements alone, and distinguishing Working Papers from Information Papers remains informative for Resolution predictions using only preceding-meeting papers. Formal continuity, therefore, coexists with informative contributions, making the ongoing work relevant even when agreement subjects recur.

Differences in participants' portfolios raise questions about who sustains the shared agenda as concerns change. Producing information and proposing responses can shape which issues enter discussion, a form of agenda-setting that precedes the adoption of agreements \cite{BachrachBaratz1962}. New specializations arising in concerns closely linked to earlier portfolios make earlier specializations relevant to participation, because participants start from different combinations of work. The network provides a basis for examining whether these differences reflect a division of responsibility or constraints on participation. Following this work alongside agreements also supports assessing governance through the processes that sustain its agenda as well as its adopted outputs \cite{GutnerThompson2010}.

These inferences concern documented participation and relative emphasis. RPA does not directly identify preferences, bargaining power, or influence. Expertise, mandates, reporting duties, and resources may help explain portfolio development, but the records do not distinguish these mechanisms or establish exclusion. The prediction comparisons concern subject shares, and their gains depend on how agreement history is represented. The additional predictive contribution of attention to connected subjects is small if existent, and model choices were informed by the evaluation meetings (\cref{SI-sec:document-type-prediction}). These analyses, therefore, remain retrospective. Implementation and environmental outcomes require evidence beyond the submission and adoption records.

Antarctica's role in global climate and sea level makes its governance consequential far beyond the continent. Understanding how this global commons is governed requires following the work through which participants sustain and develop a shared agenda, alongside the agreements they adopt. Connections among concerns make this distributed work visible, allowing governance to be studied through the evolving contributions of its participants.

\subsection*{CREdiT author statement}
\textbf{Casper van Elteren:} Conceptualization; Methodology; Software; Formal analysis; Investigation; Data curation; Visualization; Writing – original draft; Writing – review \& editing. \textbf{Fiona Lippert:} Conceptualization; Writing – review \& editing. \textbf{Vítor V. Vasconcelos:} Conceptualization; Writing – review \& editing. \textbf{Zach Carter:} Writing – review \& editing. \textbf{Maria Kleshnina:} Supervision; Funding acquisition; Writing – review \& editing. \textbf{Michael Bode:} Writing – review \& editing.

\subsection*{Declaration of interests}
All authors declare no competing interests.

\subsection*{Declaration of use of AI}
We used large language models to assist with manuscript editing to fix grammar mistakes and improve clarity. The output was carefully checked, and all edits were approved by the authors.

\subsection*{Ethics declaration}
This study analyses publicly available institutional documents and does not involve human participants, personal data, or animal research.

\subsection*{Resource availability}
\noindent\textbf{Lead contact:} Casper van Elteren (casper.vanelteren@qut.edu.au).\\
\textbf{Materials availability:} This study analyses existing documentary records and generated no new materials.\\
\textbf{Data and code availability:} The June 2026 analysis archive is available at Zenodo (\url{https://doi.org/10.5281/zenodo.20821775}). We have prepared a self-contained review archive with the reconstructed many-to-many paper panel, the pinned corpus of 277 official Measure body texts, formal-output inputs, analysis scripts, and derived tables. We will deposit the same materials as a new version of the Zenodo record before publication. Analysis scripts use Python 3.13 with NumPy, pandas, and SciPy; figures use UltraPlot \cite{vanElteren2025}. We maintain development code at \url{https://github.com/cvanelteren/theSpaceOfConcerns}.

\subsection*{Funding}
The Australian Research Council Special Research Initiative Securing Antarctica's Environmental Future (SR200100005) and an Australian Research Council DECRA Fellowship to Maria Kleshnina (DE250101223) supported this work.

\subsection*{Acknowledgments}
C.v.E. thanks Larissa Lubiana Botelho for discussions and comments on the manuscript and the Antarctic Treaty System.

\printbibliography
\clearpage

\sloppy
\twocolumn[{
\centering
\Large\textbf{Supplementary Information}\par\medskip
\large\thetitle\par\bigskip
}]

\appendix
\setcounter{figure}{0}
\setcounter{table}{0}
\renewcommand{\thefigure}{S\arabic{figure}}
\renewcommand{\thetable}{S\arabic{table}}

The Supplementary Information provides the definitions and analyses supporting the main text.

\begin{itemize}
 \item \textbf{Data and Methods.} Archive coverage, definitions of attention and specialization, construction and checks of the concern network, and methods for studying portfolio development and agreement subjects (\cref{SI-sec:si-methods}).
 \item \textbf{Supplementary Results.} Portfolio positioning, bridge-rank stability, concern profiles, prediction results and checks of current and earlier specifications, formal citations among Measures, comparisons across participant groups and cohorts, matched portfolio tests, and the Madrid Protocol boundary test (\cref{SI-sec:si-results}).
\end{itemize}

\section{Data and Methods}\label{SI-sec:si-methods}
\subsection{Data and archive coverage}
We analyse working papers and information documents submitted from 1961 to 2025. ``ATS'' refers to the broader governance system anchored in the Antarctic Treaty, but our corpus comes only from the Treaty-centred archive of Antarctic Treaty Consultative Meetings (ATCMs). The Secretariat's taxonomy also includes concerns handled by the Committee for Environmental Protection because these matters reach the ATCM agenda. Meeting volume rises partly for institutional reasons: ATCMs became annual in the 1990s, and the Committee has normally met annually since 1998. Conference Room Papers lack systematic availability, so we exclude them. The corpus therefore observes formal documentary preparation, not the informal negotiation record.

Working Papers address matters requiring discussion and action and can be submitted to the ATCM by Consultative Parties and observers. Information Papers provide supporting material on issues relevant to the meeting and can also be submitted by non-Consultative Parties and invited experts \cite{ATCMRules2015,ATSDelegatesManual2025,Gardiner2025}.

The official archive supplies thematic search categories, submitters, and document types \cite{Yaryab2024}. We retain Secretariat papers, so the Secretariat appears both as maintainer of the database categories and as one of the 66 participants. Its papers enter on the same terms as those of other participants.

We collected the data on 6 February 2026 from the ATS website (\url{https://www.ats.aq/devAS/Meetings/DocDatabase?lang=e}) with Carlo Hamalainen's antarctic-database-go collector (\url{https://github.com/carlohamalainen/antarctic-database-go}). The collector stores HTTP responses, source documents, and processing status in SQLite. These cached records permit deterministic reruns even if the website changes.

We do not derive the categories from paper text. They are the harmonized cross-meeting categories under which the Secretariat-maintained database returns each paper; we refer to them as concerns. A paper can appear under several categories, although the returned paper payload does not itself carry a category field. The query responses also record no category-assignment date. The concern network therefore describes co-specialization within an institutional taxonomy, not semantic similarity. We will include the processed many-to-many panel and reconstruction code in the next Zenodo version.

We reconstructed paper--category relations from the collector's cached category-query responses, then deduplicated attachment-linked records by \texttt{paper\_id}. This produced 6,591 unique papers. We removed 18 papers (0.3\%, all after 2010) returned only under the placeholders \texttt{ALL} or \texttt{Other}, leaving 6,573 papers and 7,895 substantive paper--category relations. Of the retained papers, 5,345 have one category and 1,228 have two to five. We divide a paper with $k$ categories equally among them, so it contributes $1/k$ to each label. Papers with multiple sponsors are then expanded to one row per participant; each sponsor receives the paper's same fractional category weights. The extracted and reconstructed dataset contains the following columns.

A separate snapshot from 29 January 2024 contains the same 45 categories. All 6,230 matched papers with a nonmissing snapshot category retain that category in the 2026 reconstruction. The snapshot stores only one category per paper, does not date that assignment, and does not cover ATCMs 46--47. It supports label stability but cannot establish when every multi-label membership became available.

\begin{figure*}[!htbp]
\centering
\includegraphics[width=\textwidth]{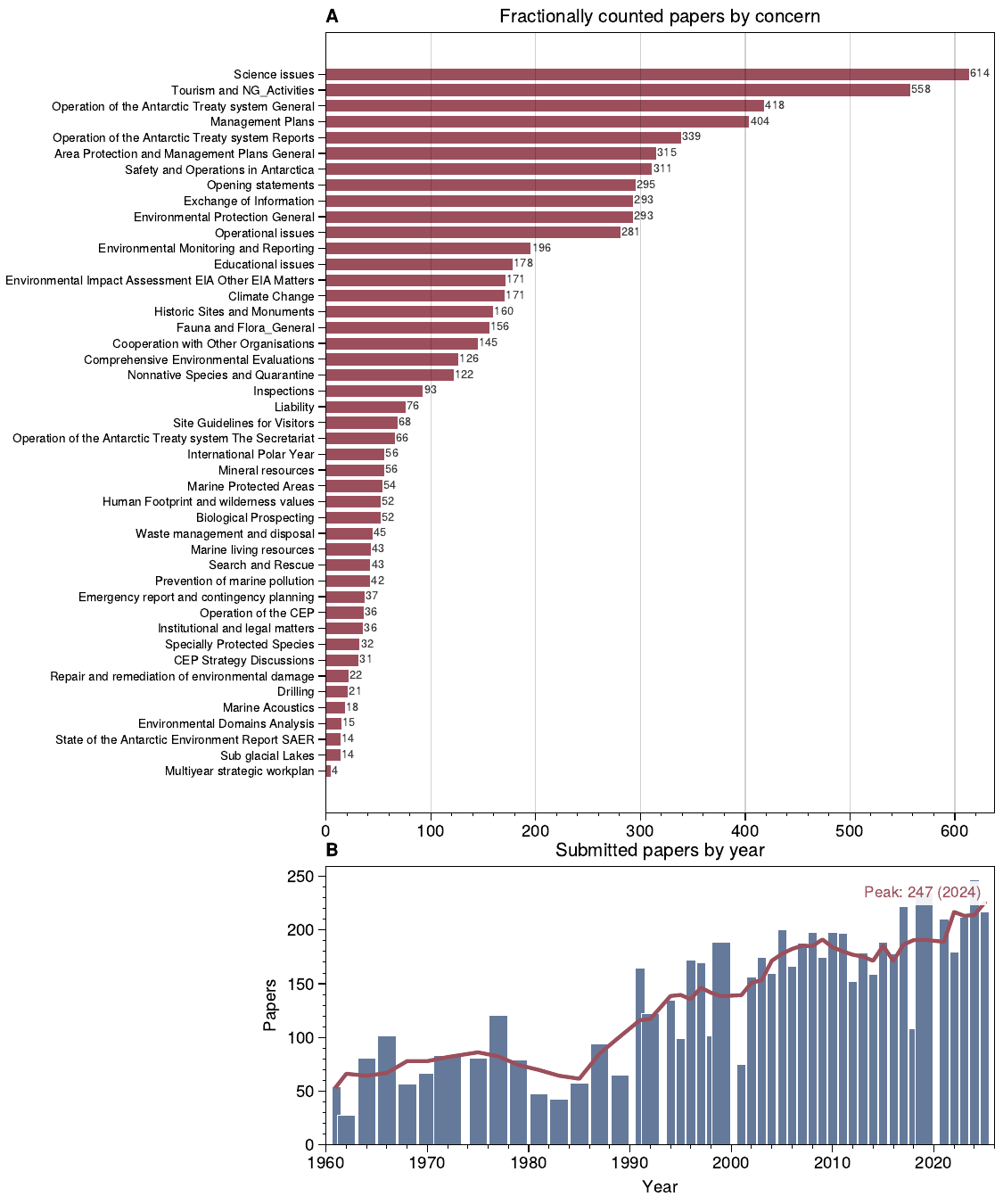}
\caption{\textbf{Paper submissions rise after 1990 and remain uneven across concerns.} (a) Fractional counts for 6,573 working and information papers by Secretariat concern, 1961--2025. A paper returned under $k$ concerns contributes $1/k$ to each. (b) Unique papers submitted each year. The line is the centred five-year average.}
\label{SI-fig:topic-submission-overview}
\end{figure*}

{\small
\vspace{1em}
\begin{tabular}{@{}p{0.45\columnwidth} p{0.45\columnwidth}@{}}
\texttt{meeting\_year} & \texttt{paper\_id} \\
\texttt{meeting\_type} & \texttt{party\_type} \\
\texttt{meeting\_number} & \texttt{paper\_name} \\
\texttt{meeting\_name} & \texttt{paper\_number} \\
\texttt{party} & \texttt{paper\_revision} \\
\texttt{category} & \texttt{paper\_language} \\
\texttt{page\_url} & \texttt{paper\_url} \\
\texttt{page\_nr} & \texttt{exists} \\
\texttt{payload\_json} & \texttt{agendas} \\
\texttt{attachment\_id} & \texttt{parties} \\
\texttt{attachment\_name} &\texttt{attachment\_language} \\
\texttt{attachment\_url} & \texttt{attachment\_exists} \\
\vspace{1em}
\end{tabular}\par}
The June 2026 archive is available at \url{https://doi.org/10.5281/zenodo.20821775}; we will add the reconstructed many-to-many panel in the next version. \Cref{SI-fig:topic-submission-overview}a summarizes the resulting distribution across concerns, while panel b shows annual submission volume.

The unit is the \emph{participant}, defined by submitting papers rather than by legal status under the Treaty. We include Consultative Parties, Non-Consultative Parties, Observers, invited Expert organizations, and the Secretariat because all appear as submitters in this documentary arena. Only Consultative Parties adopt agreements. The participant types are therefore not institutionally equivalent, and aggregated estimates describe documentary activity rather than formal authority \cite{Sampaio2022}. Sixty-six participants appear in the corpus. The extracted field \texttt{party\_type} contains Working Paper and Information Paper codes, not Treaty status. Participant-role information is separate and can change over time. The cohort comparison retains 43 participants, while the states-only locality check retains 42 states. Conference Room Papers do not enter any analysis.

\subsection{Building the concern network}\label{SI-sec:map-construction}
We adapt the product-space construction of Hidalgo and colleagues \cite{Hidalgo2007}, which links products through the countries that specialize in them. In our setting, participants replace countries and governance concerns replace products, providing the basis for the specialization and proximity measures defined below.

Counts and co-specialization answer different questions. Counts show where attention is common. Co-specialization shows which concerns receive a disproportionate share of the same participants' papers. A concern pursued by nearly everyone can be prominent in counts but uninformative for co-specialization because no participant stands out. A small technical group can produce the opposite pattern. Our network records which concerns share specializing participants, not the overall importance of a concern.

Let $x(a,i)$ denote the fractionally counted papers submitted by participant $a$ on concern $i$. If the archive returns paper $p$ under $k_p$ substantive categories, it contributes $1/k_p$ to each; a co-sponsored paper contributes those weights to every listed participant. Participants include Consultative Parties, Non-Consultative Parties, Observers, invited Expert organizations, and the ATS Secretariat whenever they appear as submitters. We define the normalised activity share
\begin{equation}\label{SI-eq:activity_share}
\pi(a,i) = \frac{x(a,i)}{\sum_j x(a,j)},
\end{equation}

and the system-wide importance of concern $i$ as

\begin{equation}\label{SI-eq:topic_importance}
Z_i = \frac{\sum_c x(c,i)}{\sum_{d,j} x(d,j)}.
\end{equation}

The Revealed Policy Advantage (RPA) index is then
\begin{equation}\label{SI-eq:rpa}
 \rho(a,i) = \frac{\pi(a,i)}{Z_i}.
\end{equation}
We call this ratio \emph{revealed policy advantage} rather than the usual \emph{revealed comparative advantage} (RCA) because the units are policy concerns rather than exported products. The mathematics does not change, but the interpretation is narrower. Submissions have not passed a competitive market test. Participants may submit to inform, comply, signal, or seek action. RPA records the concerns receiving a disproportionate share of a participant's papers. It cannot separate intent from mandate, expertise, reporting duties, or document-production capacity, and it cannot observe priorities that produce no paper. Scientific capacity may also enter the measure because Article IX(2) ties consultative status to the scale of scientific research activity. Nor do concerns have an equivalent of product complexity or economic value. Co-specialization remains useful because it records how participants allocate finite documentary effort across the agenda.

We measure co-specialization by how strongly each pair of concerns shares specializing participants. We call this value proximity; it gives the link weights of the network of concerns:
\begin{equation}\label{SI-eq:proximity}
\phi(i,j)
=
\min\!\left\{
\begin{aligned}
 &\Pr\!\bigl(\rho(a,i) \ge 1 \mid \rho(a,j) \ge 1\bigr),\\
 &\Pr\!\bigl(\rho(a,j) \ge 1 \mid \rho(a,i) \ge 1\bigr)
\end{aligned}
\right\}.
\end{equation}
with
\begin{equation}
 \Pr(\rho(a,j)\ge1 \mid \rho(a,i)\ge1) =\frac{\sum_a \mathbf{1}[\rho(a,i)\ge1 \land \rho(a,j)\ge1]}
 {\sum_a \mathbf{1}[\rho(a,i)\ge1]},
\end{equation}
This produces a weighted network of relations among concerns. Conditional probability allows two uncommon concerns to be close when specialization in one reliably accompanies specialization in the other; a joint probability would shrink mechanically with rarity. Taking the lower of the two directions prevents a niche concern from appearing close to a broad concern merely because the niche usually occurs alongside it. The resulting score is symmetric. We estimate it across participants in the relevant window. With 66 participants, individual relations rest on modest counts: the strongest, between Marine living resources and Mineral resources, rests on 6 participants. The network is therefore a coarse aggregate rather than a high-resolution estimate of every pair. We construct the full-record network and rebuild time-specific networks within each window.

\subsection{Participant-group panels}\label{SI-sec:group-specialization}
For \cref{fig:the_space_of_concerns}c--e, we calculate each participant's RPA from their full record, using all 66 participants as the common reference. For concern $i$ and group $g$, the displayed proportion is $\sum_{a\in g}\mathbf{1}[\rho(a,i)\ge1]/n_g$. The corpus contains 29 Consultative Parties, 18 Non-Consultative Parties, and 19 organizations and institutional bodies, including the Secretariat. These are represented participants, not total Treaty membership; parties absent from the corpus are not assigned zero specialization. We retain one classification per participant across the full record, including papers predating consultative status. Organizations and institutional bodies are grouped for display without implying equivalent formal roles. The numerator counts individual specializations rather than computing RPA from pooled group contributions. Each participant contributes at most one count per concern, irrespective of paper volume or how far RPA exceeds 1. The group panels describe specialization prevalence rather than absolute activity or underlying intent.

\begin{figure*}[!htbp]
\centering
\includegraphics[width=\textwidth]{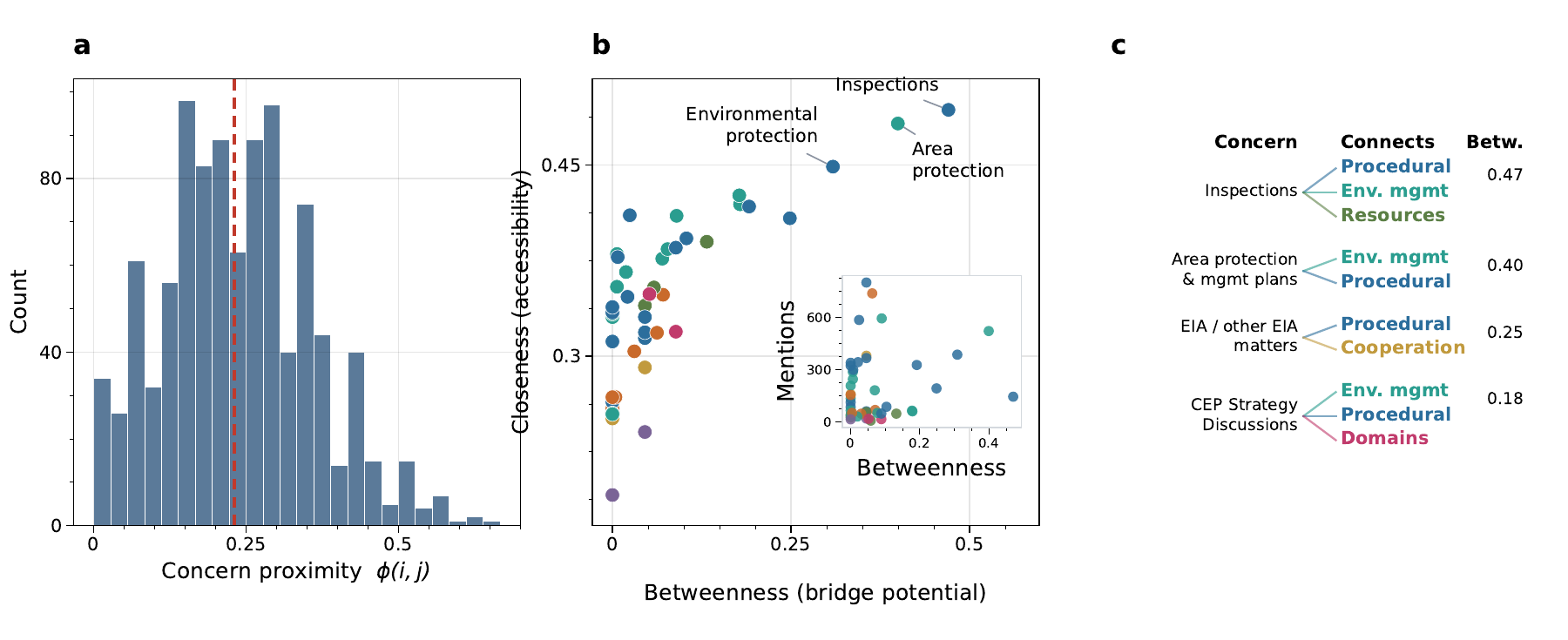}
\caption{\textbf{Concern proximity and network position describe distinct aspects of the shared agenda.} (a) Distribution of proximity scores for all 990 concern pairs, including the 34 zero-proximity pairs; the dashed line marks the median. (b) Closeness and betweenness on the displayed backbone, with fractional paper contributions against betweenness inset. Colours follow the concern regions in \cref{fig:the_space_of_concerns}. (c) Four concerns with links to multiple regions, ranked by betweenness, and the regions they connect. Centralities use proximity-derived distances on the same maximum spanning tree plus bootstrap-supported links shown in \cref{fig:the_space_of_concerns}b.}
\label{SI-fig:concern-network-diagnostics}
\end{figure*}

\subsection{Can marginal counts reconstruct the concern network?}\label{SI-sec:counts-versus-relations}
To test whether proximity is only a reformulation of prevalence, we treat each of the 990 concern pairs as one observation and use full-record proximity $\phi(i,j)$ as the outcome. For each pair, the paper-count model contains the sum, absolute difference, and minimum of the two concerns' log-transformed marginal paper volumes. The participant-prevalence model uses the corresponding functions of the number of participants with $\mathrm{RPA}\ge1$. A combined model includes both sets. Each specification is an ordinary least-squares linear regression with an intercept. These are reconstruction diagnostics, not causal models.

\begin{figure}[!htbp]
\centering
\includegraphics[width=\linewidth]{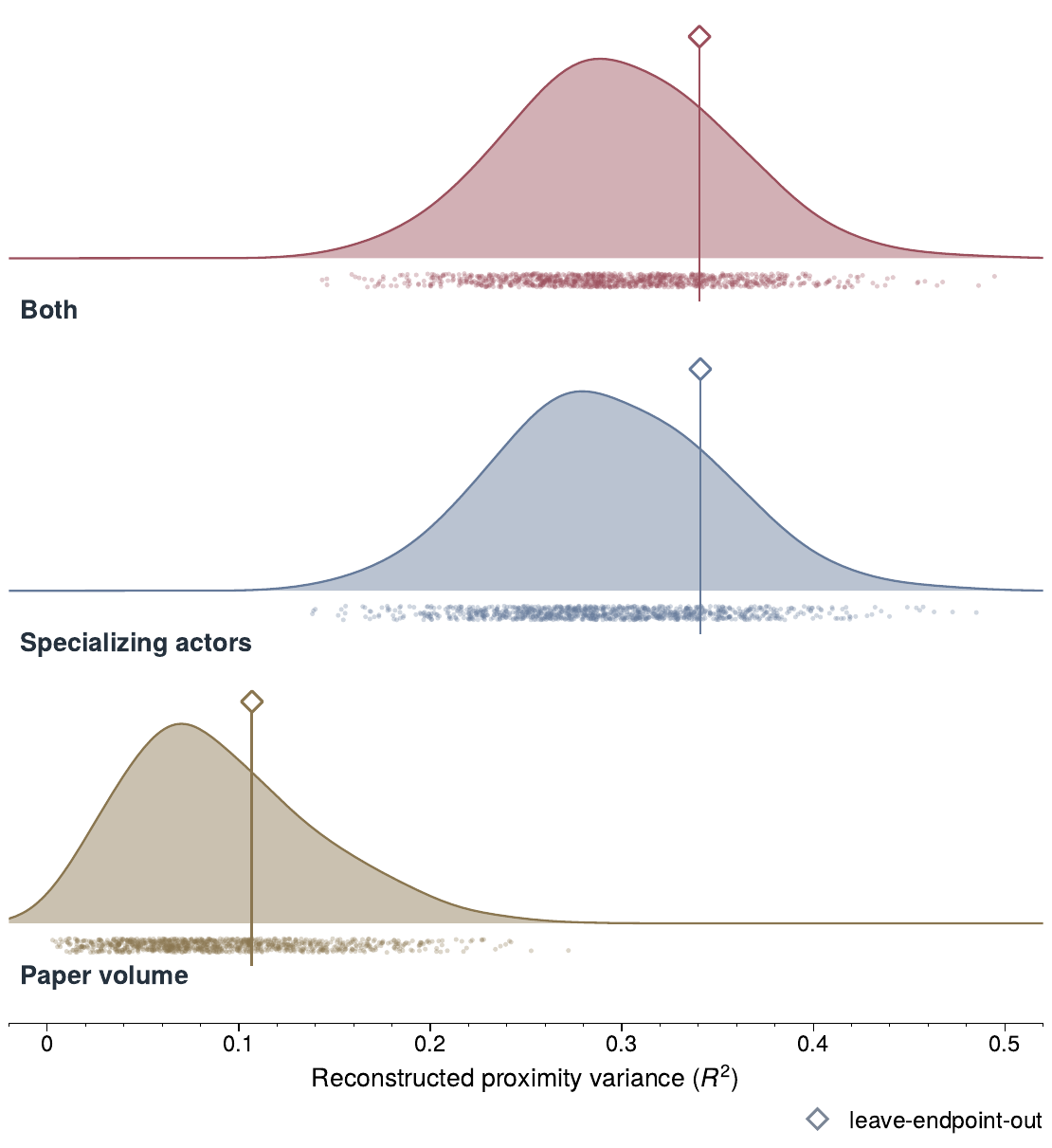}
\caption{\textbf{Marginal counts reconstruct only part of co-specialization.} Half-violins and rugs show pooled $R^2$ across 1,000 participant bootstraps and 990 concern pairs for models using paper volume, specializing-participant counts, or both. Open diamonds mark leave-endpoint-out $R^2$, which predicts pairs containing a concern excluded from fitting.
}
\label{SI-fig:the-space-of-concerns-reconstruction-appendix}
\end{figure}

We assess generalization with a leave-endpoint-out procedure. For each concern, we omit every pair containing it, fit the model to the remaining pairs, and predict the omitted relations. We predict each pair twice, using one fit that excludes each endpoint, and average the two predictions before calculating $R^2$. We then resample the 66 complete participant histories 1,000 times and rebuild paper volume, RPA assignments, the full proximity matrix, and every model in each draw. Finally, 9,999 quadratic assignment procedure (QAP) label permutations randomly reassign the observed marginal attributes to concerns while leaving the matrix fixed. This preserves the marginal distributions while testing whether their observed alignment with the concern network exceeds arbitrary assignments.

Paper volume explains $12.8\%$ of pairwise proximity variation in the pooled fit and $10.7\%$ when predicting relations involving a concern omitted from model fitting (\cref{SI-fig:the-space-of-concerns-reconstruction-appendix,SI-tab:count-reconstruction}). Specialization prevalence explains $35.3\%$ under that fit and $34.1\%$ under the same test. The combined model reaches $35.9\%$ across all pairs but remains at $34.1\%$ when the procedure omits concerns. Across participant bootstraps, the additional all-pairs variance explained by paper volume after participant prevalence has a median of $0.005$ and a central 95\% interval of $[0.0003,0.0173]$. All three observed pooled $R^2$ values exceed every corresponding QAP draw ($p=0.0001$). Marginal prevalence therefore aligns with the concern network, as expected for a score constructed from specialization, but it does not reconstruct most pairwise variation. \url{scripts/counts_vs_relations.py} reproduces the analysis.

\begin{table*}[!htbp]
\centering
\footnotesize
\caption{\textbf{Marginal counts recover only part of pairwise co-specialization.} Pooled $R^2$ uses all 990 concern pairs. Leave-endpoint-out $R^2$ predicts pairs containing an excluded concern. Participant-bootstrap entries are medians with central 95\% intervals from 1,000 resamples. Quadratic assignment procedure (QAP) $p$ values use 9,999 random assignments of the marginal attributes.}
\label{SI-tab:count-reconstruction}
\begin{tabular}{@{}lcccc@{}}
\hline
Model & Pooled $R^2$ & Leave-endpoint-out $R^2$ & Participant-bootstrap $R^2$ & QAP $p$ \\
\hline
Paper volume & 0.128 & 0.107 & 0.083 $[0.016,0.199]$ & 0.0001 \\
Specializing participants & 0.353 & 0.341 & 0.288 $[0.182,0.409]$ & 0.0001 \\
Both & 0.359 & 0.341 & 0.295 $[0.188,0.413]$ & 0.0001 \\
\hline
\end{tabular}
\end{table*}

\subsection{Testing whether new specializations arise in closely linked concerns}\label{SI-sec:hazard-model}
We test whether new specializations arise in concerns closely linked to a participant's earlier portfolio, a tendency we call \emph{locality}, by comparing adjacent rolling five-meeting windows. A concern enters the comparison only when the participant was not already specialized in it. For participant $a$, concern $i$, and transition from $t-1$ to $t$, this condition is
\[
\text{at-risk}_{a i t}=1 \iff \rho_{a i, t-1} < 1.
\]
The binary outcome is then
\[
y_{a i t}=1 \iff \rho_{a i, t} \ge 1 \ \text{and}\ \rho_{a i, t-1} < 1.
\]
Let the previously active support be
\[
S_{a,t-1}=\{j:\rho_{aj,t-1}\ge 1\},
\]
let $A_{t-1}$ be the concerns that have appeared in the archive before the transition, and let the available set be
\[
R_{at}=\{i\in A_{t-1}:\rho_{ai,t-1}<1\},
\]
and let $m_{at}=\sum_{i\in R_{at}} y_{ait}$ denote the number of new specializations in a participant-period. Our main locality specification is then a participant-period conditional logit,
\begin{dmath}
\Pr(\mathbf{y}_{at}\mid m_{at}, X_{at})
\propto
\exp\!\left(\sum_{i\in R_{at}}y_{ait}\bigl(\beta_d d_{a i, t-1}\!+\! \beta_p \mathrm{pop}_{i,t-1}
\bigr)
\right),
\end{dmath}
where
\[
d_{a i, t-1}=1-\max_{j\in S_{a,t-1}}\phi(i,j)
\]
is the distance from candidate concern $i$ to the participant's earlier portfolio and $\mathrm{pop}_{i,t-1}$ is its prior popularity (the share of participants with $\mathrm{RPA} \ge 1$). The model holds constant how many concerns cross the threshold in that period. It therefore compares which available concerns cross for the same participant at the same time, rather than comparing participants with different general rates of expansion. Periods in which a participant has no earlier portfolio or no new specialization do not enter the model. Differences in portfolio size are constant within each participant-period comparison. The network, portfolios, and new specializations all use $\mathrm{RPA}\ge1$.

These restrictions leave 968 participant-period comparisons and 30,957 participant-period--concern options. The distance coefficient is $-2.02$ (standard error [SE] $0.22$), or an odds ratio of $0.817$ per $+0.1$ distance (95\% model-based confidence interval [CI] $[0.783,0.853]$). The specification is primary because neither the network nor the choice set uses information first observed during the transition. A second outcome requires no paper on the concern in the earlier window and at least one in the next. It gives an odds ratio of $0.874$ ($[0.834,0.916]$), showing that the local association is not only a consequence of the relative specialization threshold. \url{scripts/compare_major_results_category_treatments.py} reproduces both estimates.

We construct the example trajectories in Supplementary \cref{SI-fig:actor-continuous-trajectories} on a continuous axis. We compute RPA in rolling five-meeting windows and locate each participant on the horizontal axis of the fixed full-record network as the RPA-weighted mean of its concern coordinates. We omit windows with fewer than three concerns at $\mathrm{RPA}\ge1$. Australia connects this continuous summary to the snapshots in \cref{fig:local-specialization}, while the Netherlands and Ukraine provide two further examples. The trajectories illustrate movement rather than testing locality. \texttt{figS22\_actor\_trajectories.py} reproduces the figure.

\begin{figure*}[!htbp]
\centering
\includegraphics[width=0.92\textwidth]{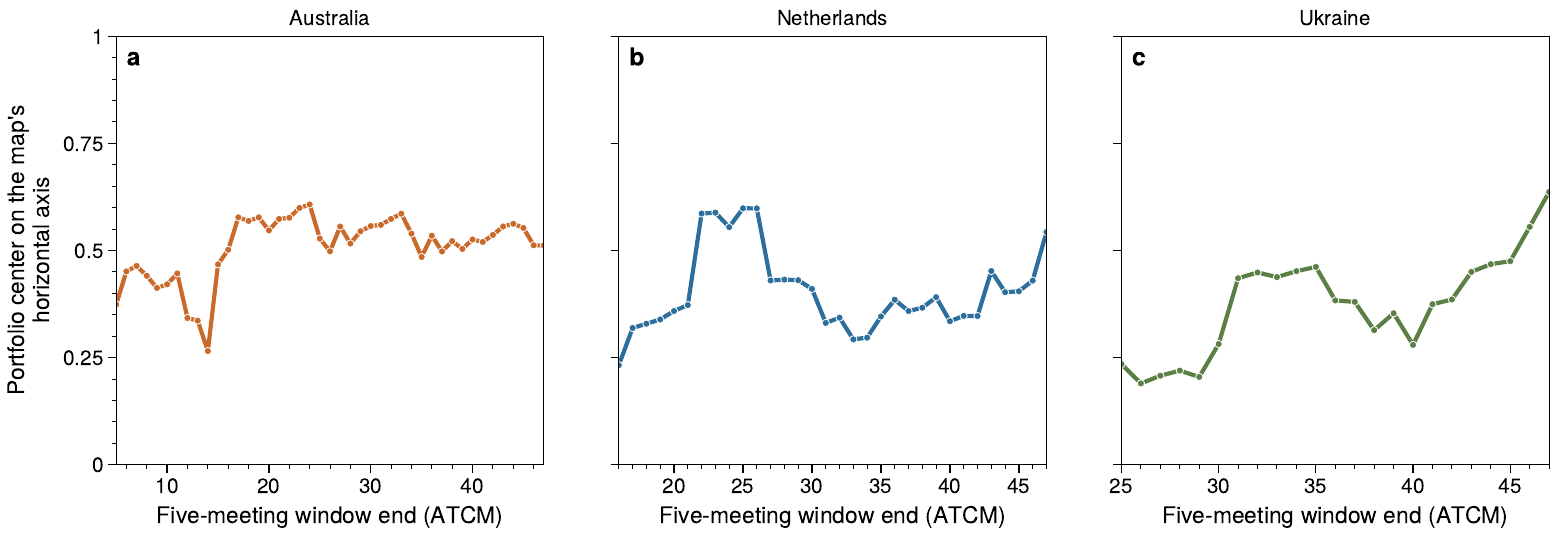}
\caption{\textbf{Three participant portfolios move across the concern network.} Lines trace the revealed-policy-advantage-weighted mean horizontal coordinate of Australia, the Netherlands, and Ukraine on the fixed full-record network over rolling five-meeting windows. We omit windows with fewer than three specialized concerns. The trajectories illustrate movement only.}
\label{SI-fig:actor-continuous-trajectories}
\end{figure*}

Because the rolling windows overlap, successive choice sets for the same participant are not statistically independent. We therefore resample complete participant histories and rebuild portfolios, historical availability, the network from all earlier records, and the choice model over 500 bootstrap replicates. All 500 fits succeed. The distribution of full-rebuild bootstrap estimates has median OR $0.869$ and central 95\% percentile range $[0.820,0.918]$. After rebuilding, the median is closer to one than the original OR $0.817$; the percentile range summarizes rebuilt estimates, not the model-based confidence interval for that original estimate. The basic 95\% bootstrap confidence interval for that original estimate, calculated on the distance-coefficient scale before transformation to an OR, is $[0.727,0.814]$. The basic interval reflects the bootstrap quantiles around the original coefficient; their shift toward zero explains why its upper bound falls below the original OR. We retain $0.817$ as the primary point estimate and report the rebuilt bootstrap distribution as a sensitivity check. We also use disjoint five-meeting blocks and restrict the panel to states. The blocks reduce rolling-window overlap, but successive participant transitions still share their middle block and use networks built from nested sets of earlier records; the participant-history bootstrap addresses this dependence. The association remains negative in both checks; with disjoint blocks, the odds ratio is $0.904$ ($[0.855,0.957]$). Restricting the primary choice sets to 42 states gives OR $0.827$ (95\% model-based interval $[0.789,0.866]$), across 821 participant-periods and 25,963 concern options. This restriction retains the all-participant network and portfolios, unlike the Consultative-Party reconstruction in \cref{SI-sec:consultative-party-sensitivity}. \url{scripts/report_states_only_locality.py} reproduces this estimate. We report these checks in \cref{SI-tab:locality_estimands}. The result also remains stable across the full-record, previous-window, and earlier-record networks while holding fixed the set of options available at the time (\cref{SI-fig:hazard-space-sensitivity}). The primary distance uses raw $1-\phi$, which has bounds and a direct interpretation; $-\log\phi$ is unstable when two concerns have zero proximity. \url{scripts/bootstrap_fractional_locality.py} and \url{scripts/hazard_robustness_dependence.py} reproduce the dependence checks.

\subsection{How long specialization lasts}\label{SI-sec:retention-method}
\begin{figure}[!htbp]
\centering
\includegraphics[width=0.46\textwidth]{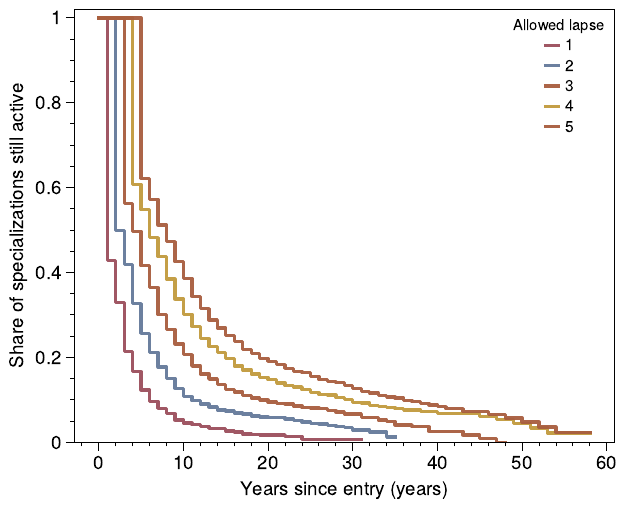}
\caption{\textbf{Estimated specialization duration depends on the allowed inactivity gap.} Kaplan--Meier curves show the share of annual specialization spells, defined by revealed policy advantage (RPA) $\ge1$, still active when failure requires more than one to five consecutive inactive years. The five-year curve remains at one through year five. Survival after ten years ranges from $0.05$ to $0.42$. After twenty years every estimate is below $0.20$. We right-censor spells active in 2025. The archive cannot distinguish a concern-specific lapse from a participant-year with no submissions.}
\label{SI-fig:retention-curves}
\end{figure}

We estimate specialization spells on 1-year windows. A spell begins when a participant first crosses $\mathrm{RPA}\ge1$ for a concern, and we follow it until inactivity exceeds an allowed gap of one to five consecutive years. We right-censor spells still active at the end of 2025 and compute a Kaplan--Meier curve for each gap definition. The allowed gap delays any failure declaration, so every curve mechanically equals one over that initial interval. \Cref{SI-fig:retention-curves} reports all five definitions on the unshifted scale of time since the spell began. The analysis does not distinguish a concern-specific lapse from years in which the participant submits nothing, so it is a descriptive sensitivity analysis rather than evidence of a retention mechanism.

\subsection{A retrospective comparison at the same career stage}\label{SI-sec:cohort-method}
The cohort analysis in \cref{SI-sec:secondary-structure-results} compares the concerns in each participant's opening portfolio with a weighted comparison. The comparison weights concerns by their share of recorded specializations during that participant's first 15 years. It scores specializing participants' portfolio size and closely linked concerns using the full-record network, so this is a retrospective description rather than a reconstruction of what participants could know at the time.

A participant is eligible if at least $\mathrm{WIN}=15$ years separate its first and last appearance in the archive; the requirement excludes participants joining after 2010 by construction, since their record cannot yet span the window. Its \emph{opening window} runs from its first recorded year to that year plus $\mathrm{WIN}-1$. Within that window we rebuild the participant--concern interaction matrix from the submissions falling inside it, compute RPA exactly as in the full-record construction, and treat a participant as specializing in a concern when $\mathrm{RPA}\ge 1$. We drop participants specializing in fewer than three concerns in their own opening window to avoid estimating opening position from one or two concerns. This leaves 14 participants joining 1961--80, 12 joining 1981--90, and 17 joining 1991--2010.

The comparison asks whether the participants specializing in those concerns have broader or narrower portfolios across the full record than expected, and whether those concerns have more or fewer closely linked concerns in the full-record network. The weighted comparison weights concerns by their share of recorded specialization during each participant's opening window. The 1991--2010 cohort sits at $0.929$ of this comparison for specializing participants' full-record portfolio size (95\% participant-bootstrap interval $[0.889,0.966]$) and $0.883$ for closely linked concerns in the full-record network ($[0.798,0.974]$). The two earlier cohorts sit near parity on both measures. Each of 400 bootstrap replicates resamples the 66 participants and rebuilds opening portfolios and availability weights while retaining the fixed concern scores. \url{scripts/cohort_partition_free.py} reproduces the comparison.

\begin{table*}[!htbp]
\centering
\scriptsize
\caption{\textbf{Retrospective first-15-year concern positions differ across entrant cohorts.} Ratios compare opening portfolios with a weighted comparison weighted by recorded specializations during the same first-15-year period. A value of 1 marks parity with that comparison. Specializing participants' portfolio size is the mean full-record portfolio size of participants specializing in each concern. Closely linked concerns are those linked to it in the full-record network with $\phi\ge0.30$. Intervals are participant-bootstrap 95\% intervals from 400 resamples. $n$ counts participants with a 15-year record and at least three opening-window specializations.}
\label{SI-tab:cohort-ratios}
\begin{tabular}{@{}llrrr@{}}
\hline
\textbf{Cohort} & \textbf{Scale} & \textbf{Ratio} & \textbf{95\% CI} & \textbf{$n$} \\
\hline
1961--80 & Specializing participants' portfolio size & 1.00 & [0.978, 1.016] & 14 \\
1961--80 & Closely linked concerns & 0.99 & [0.962, 1.021] & 14 \\
1981--90 & Specializing participants' portfolio size & 0.98 & [0.952, 1.012] & 12 \\
1981--90 & Closely linked concerns & 1.04 & [0.975, 1.133] & 12 \\
1991--2010 & Specializing participants' portfolio size & 0.93 & [0.889, 0.966] & 17 \\
1991--2010 & Closely linked concerns & 0.88 & [0.798, 0.974] & 17 \\
\hline
\end{tabular}
\end{table*}

\subsection{Checks on locality}\label{SI-sec:substantive-sensitivity}
This subsection tests whether locality depends on the temporal construction of the network, the definition of new specialization, self-contribution to the proximity matrix, or full rather than fractional credit for co-sponsored papers. A separate subsection below asks whether a small number of participants dominates the recovered network.

\begin{figure*}
 \centering
 \includegraphics[width=\textwidth]{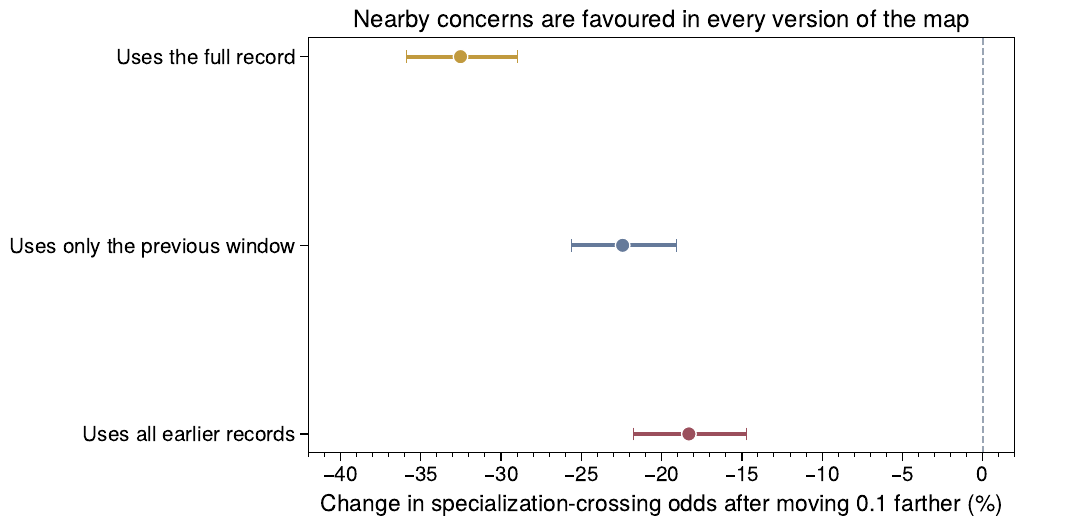}
\caption{\textbf{Locality persists across three temporal network definitions.} Across 968 participant-periods, odds ratios give the change in odds of crossing $\mathrm{RPA}\ge1$ for a $0.1$ increase in distance $d=1-\max\phi$ from the earlier portfolio. Values below one indicate locality. Bars are model-based 95\% intervals. Networks use the full record, the previous five-meeting window, or all earlier records. The earlier-record network is primary because it uses only information available before each transition. Models compare options available to the same participant at the same time and adjust for prior popularity.}
 \label{SI-fig:hazard-space-sensitivity}
\end{figure*}

We first change how we build the concern network while holding fixed the set of options available at the time. \Cref{SI-fig:hazard-space-sensitivity} compares the full-record, previous-window, and earlier-record networks. In each case, distance is $d=1-\max\phi$. Moving $0.1$ farther gives odds ratios of $0.67$, $0.78$, and $0.82$, respectively. Each model adjusts for prior concern popularity when estimating the distance association. The earlier-record network is primary because it uses no later co-specialization information. We retain raw $1-\phi$ because its values lie between zero and one; taking its logarithm is unstable when two concerns have a score of zero.

One check changes the definition of a new specialization rather than the network. A participant can cross the RPA threshold because a concern becomes less common system-wide even when its own paper count does not rise. We therefore require that the participant filed no paper on the concern in the earlier window and at least one in the next. In the corresponding participant-period conditional model, moving $0.1$ farther gives an odds ratio of $0.874$ ($[0.834,0.916]$), close to the primary direction. \url{scripts/compare_major_results_category_treatments.py} reproduces the check.

Because each participant's submission history contributes to the scores used to compute its distances, the locality test could contain a self-contribution. We therefore rebuild each participant's network from all earlier records without that participant while holding the temporally ordered outcome panel fixed. We also divide each paper equally among its sponsors, in addition to dividing it among its categories. Both checks preserve the negative distance association. The local result based only on earlier information therefore does not depend on a participant contributing to its own network or on full co-sponsor credit. \url{scripts/hazard_circularity_checks.py} reproduces both analyses.

\begin{figure}
 \centering
 \includegraphics[width=0.45\textwidth]{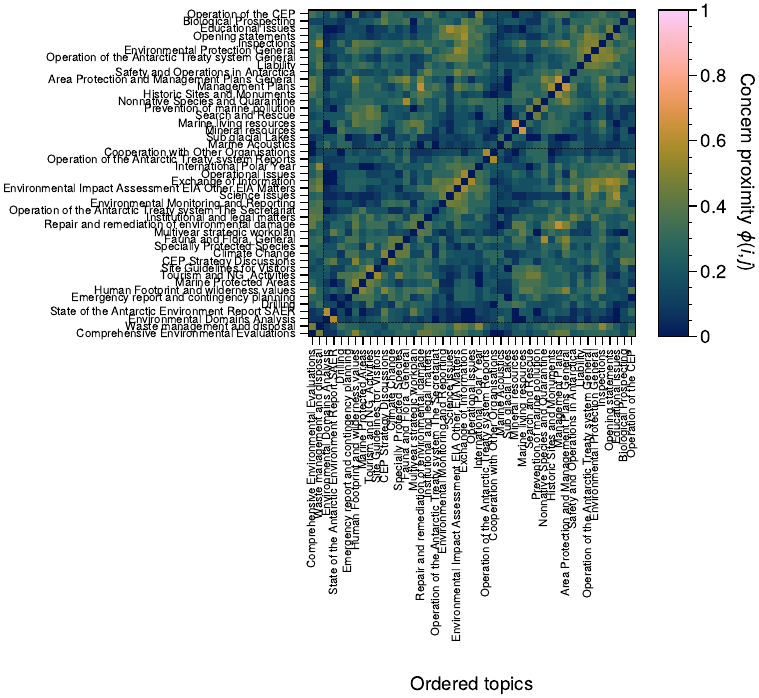}
\caption{\textbf{Co-specialization is widespread but uneven across 45 concerns.} Each cell shows the full-record score across 66 participants, defined as the lower conditional probability of specialization in one concern given specialization in the other. Darker cells indicate a stronger relation. The diagonal equals one.}
 \label{SI-fig:heatmap_concerns}
\end{figure}

Although we compute no relation from text, the recovered network could still track the wording of the official labels. We compare the proximity matrix with term frequency--inverse document frequency (TF--IDF) cosine similarity over the same 45 labels. We use word (1--2 gram, English stop words removed) and character (3--5 gram, word-boundary) analysers. A Mantel permutation with $10{,}000$ joint row-and-column shuffles preserves dependence among the 990 pairwise entries. Lexical similarity is positively but weakly associated with co-specialization, explaining $0.9\%$ of its variance for word $n$-grams ($r=0.094$, Spearman $\rho=0.066$, $p=0.008$) and $1.5\%$ for character $n$-grams ($r=0.122$, $\rho=0.157$, $p=0.002$). This test does not remove concepts already built into the taxonomy; it shows only that literal label overlap explains little of the recovered relation. The full matrix remains dense but uneven, with stronger relations concentrated within parts of the agenda (\cref{SI-fig:heatmap_concerns}). \url{scripts/semantic_baseline.py} reproduces the numbers and pair table.

\subsection{Reading the network at two levels of detail}\label{SI-sec:cpm-reading}
We use the Constant Potts Model (CPM) to group concerns with relatively strong proximity scores \cite{Traag2019}. CPM receives $\phi$ directly as the strength of a relation, not $1-\phi$ as a distance. The setting $\gamma$ controls the level of detail. CPM tends to place two concerns in the same region when their score exceeds $\gamma$ and separate them when it falls below $\gamma$. Varying $\gamma$ therefore gives progressively finer readings rather than one uniquely correct grouping.

\begin{figure*}[!htbp]
 \centering
 \includegraphics[width=\textwidth]{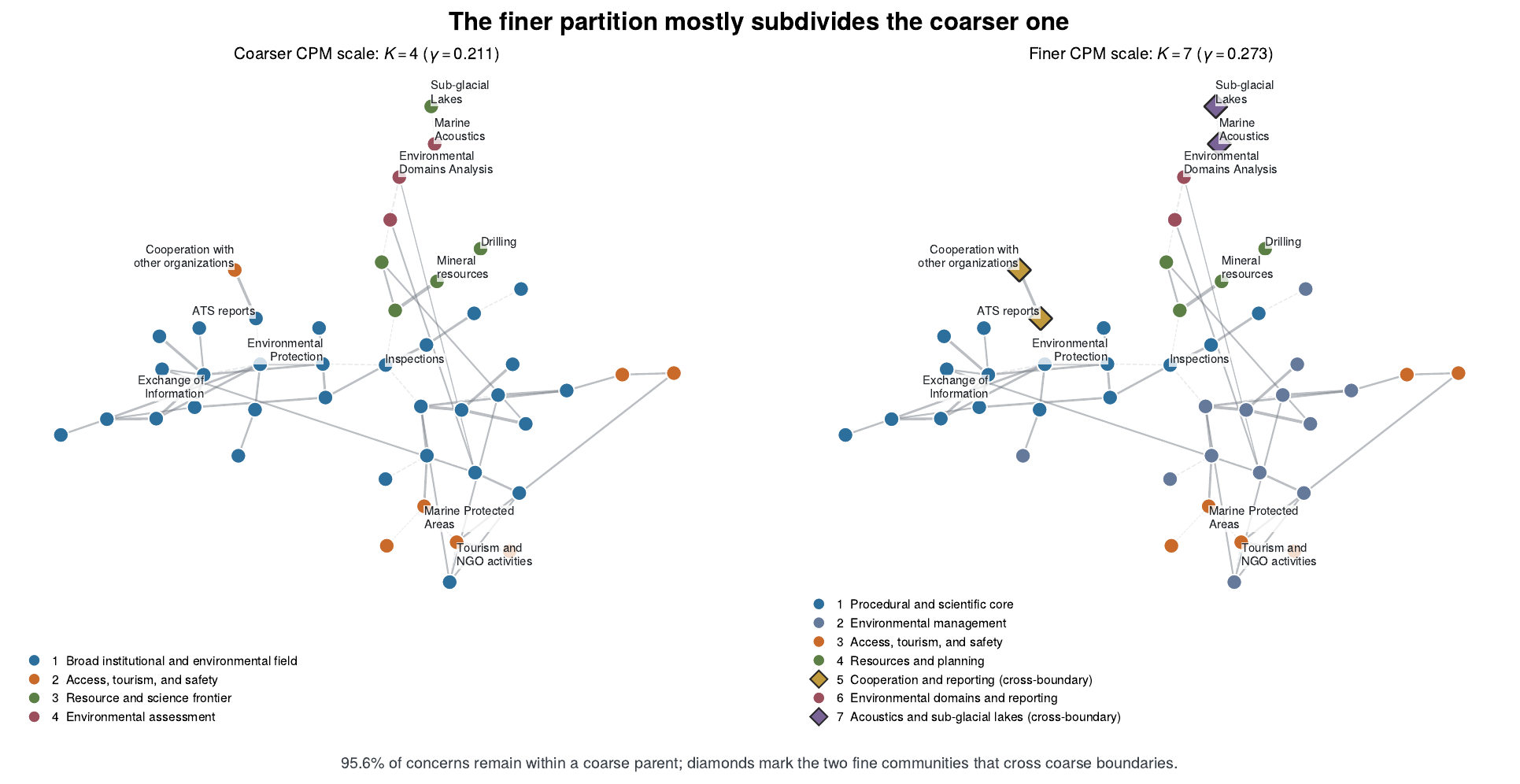}
 \caption{\textbf{Seven Constant Potts Model regions mostly subdivide the four-region structure.} The panels show $\gamma=0.211$ ($K=4$) and $\gamma=0.273$ ($K=7$) with identical concerns, relations, and coordinates. Related colours connect fine regions to coarse parents. Diamonds mark the two fine pairs that cross coarse boundaries. Overall, $95.6\%$ of concerns fall within their fine region's majority coarse parent. Neither level of detail identifies a unique grouping.}
 \label{SI-fig:cpm-partition-comparison}
\end{figure*}

The four-region reading at $\gamma=0.211$ contains groups of 30, 7, 5, and 3 concerns. The seven-region reading at $\gamma=0.273$ contains groups of 16, 13, 6, 4, 2, 2, and 2. The finer reading is mostly nested within the coarser one: 43 of 45 concerns fall within a single coarse parent, giving fine-to-coarse purity of $0.956$. Two standard measures of agreement between groupings give an adjusted Rand index of $0.478$ and normalised mutual information of $0.707$. The exceptions are Cooperation with Other Organisations and ATS reports, which form one fine pair across two coarse regions, and Marine Acoustics and Sub-glacial Lakes, which form another. \Cref{SI-fig:cpm-partition-comparison} places both groupings on identical coordinates. \url{fig_cpm_partition_comparison.py} reproduces the comparison.

\begin{figure*}[!htbp]
 \centering
 \includegraphics[width=\textwidth]{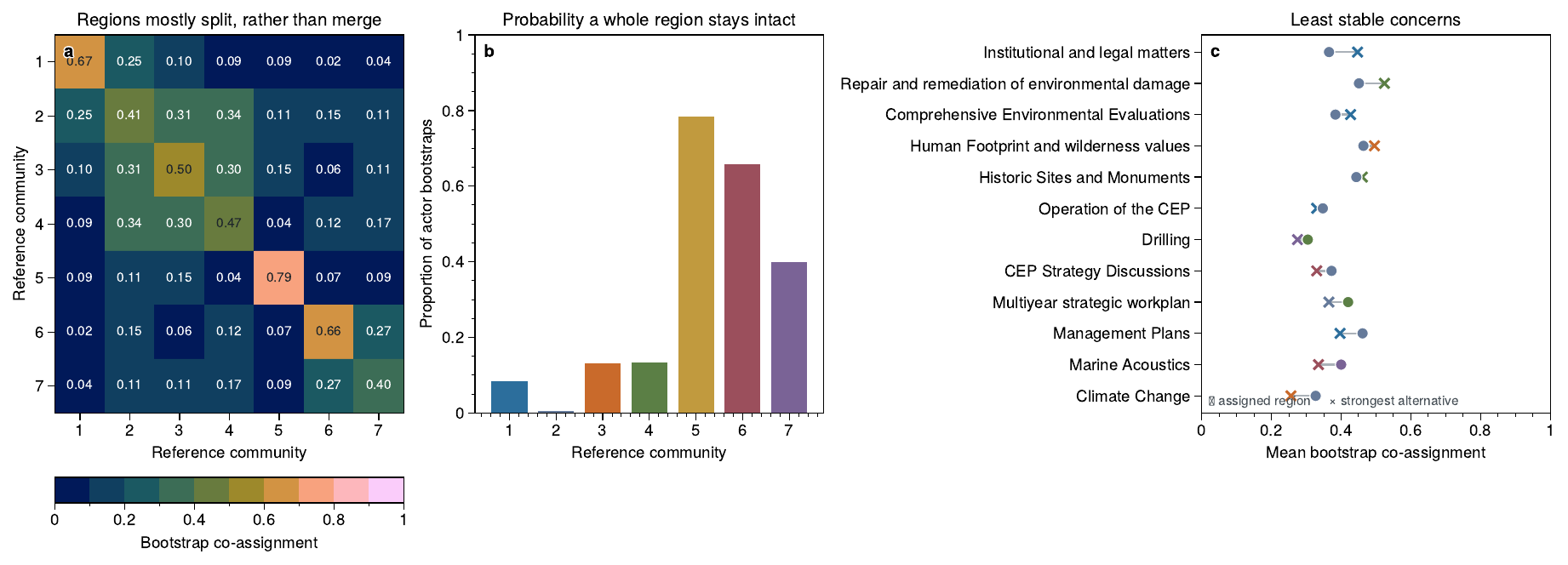}
 \caption{\textbf{Participant bootstraps split the seven displayed regions more often than they merge them.} Each of 300 resamples rebuilds RPA, proximity scores, and the Constant Potts Model grouping at $\gamma=0.273$. (a) Mean frequency with which pairs enter the same region, shown within displayed regions (diagonal) and between them (off-diagonal). (b) Probability that an entire region remains intact. (c) Twelve concerns with the smallest gap between placement with their displayed region (circle) and strongest alternative (cross).}
 \label{SI-fig:cpm-uncertainty}
\end{figure*}

To assess uncertainty, we resample participants 300 times. Within every sample, we recompute RPA, rebuild the full matrix, and refit CPM at $\gamma=0.273$. We record how often each pair enters the same region across resamples, a quantity called pairwise co-assignment frequency. This avoids matching arbitrary region numbers across bootstrap samples. The diagonal of \cref{SI-fig:cpm-uncertainty}a shows how often concerns from the same displayed region remain together; off-diagonal cells show merging across displayed regions. Internal subdivision dominates the pattern. The 16-concern procedural and scientific region has a mean within-region co-assignment frequency of $0.67$ but typically splits into three pieces. The 13-concern environmental-management region is less stable, with a mean of $0.41$. The Cooperation--Reports pair and the Environmental Domains--State of the Antarctic Environment Report (SAER) pair remain together in $79\%$ and $66\%$ of samples. Marine Acoustics and Sub-glacial Lakes remain together in $40\%$. Whole-region persistence is lower because every within-region pair must remain together (\cref{SI-fig:cpm-uncertainty}b). The smallest gaps between displayed and alternative assignments identify the least stable concern placements (\cref{SI-fig:cpm-uncertainty}c). We use pairwise co-assignment frequency as the primary comparison because whole-region persistence decreases mechanically with region size. \url{scripts/analyze_cpm_partition_uncertainty.py} reproduces the bootstrap.

\subsection{Do specialization counts explain the network?}\label{SI-sec:fixed-margin-null}
We test whether the number of concerns in which each participant specializes and the number of participants specializing in each concern explain the network. The shuffled comparison begins with the binary participant--concern matrix at $\mathrm{RPA}\ge1$. Curveball trades randomize which participants specialize in which concerns while preserving both sets of totals. We generate 1,000 shuffled networks and recompute the proximity scores, number of positive relations, overall connectivity, and weighted modularity after each draw. We optimize modularity separately in every observed and shuffled network using four independent starts of the Louvain algorithm. Empirical upper-tail probabilities include the observed statistic.

The seven displayed CPM regions come from the observed network, so holding their labels fixed in shuffled draws gives supporting evidence rather than an independent test. A second comparison identifies the best grouping separately within every draw. We also repeat the shuffling after filtering participants at paper-volume thresholds of 5, 10, 20, and 50 and portfolio-size thresholds of 3 and 5. A mixing check compares chains with 5, 20, and 50 trades per observed membership. \url{scripts/topology_fixed_margin_null.py} reproduces the comparison, and \url{output/scientific_checks/topology_curveball_mixing_summary.json} records the mixing check.

All 45 concerns belong to one connected network, and only 34 of 990 concern pairs have a proximity score of zero (\cref{SI-fig:concern-network-diagnostics}a). Shuffled comparisons preserving the number of specializations per participant and per concern reproduce most of this connectivity. The observed network nevertheless has greater weighted modularity, which measures the concentration of connections within groups: $Q=0.090$ versus a shuffled median of $0.052$ ($p=0.001$). The modest modularity supports a coarse description of local concentrations within a broadly connected agenda.

\subsection{Does any one participant determine the network?}\label{SI-sec:space-robustness}
We test whether any one participant determines the network by removing participants before rebuilding it. At each step, we remove the participant whose exclusion produces the largest root mean squared difference from the original 66-participant matrix, then repeat the procedure. This does not test a substantive behavioral claim directly; it asks whether many participants jointly produce the pattern or a small number dominate it.

\begin{figure}
 \centering
 \includegraphics[width=0.45\textwidth]{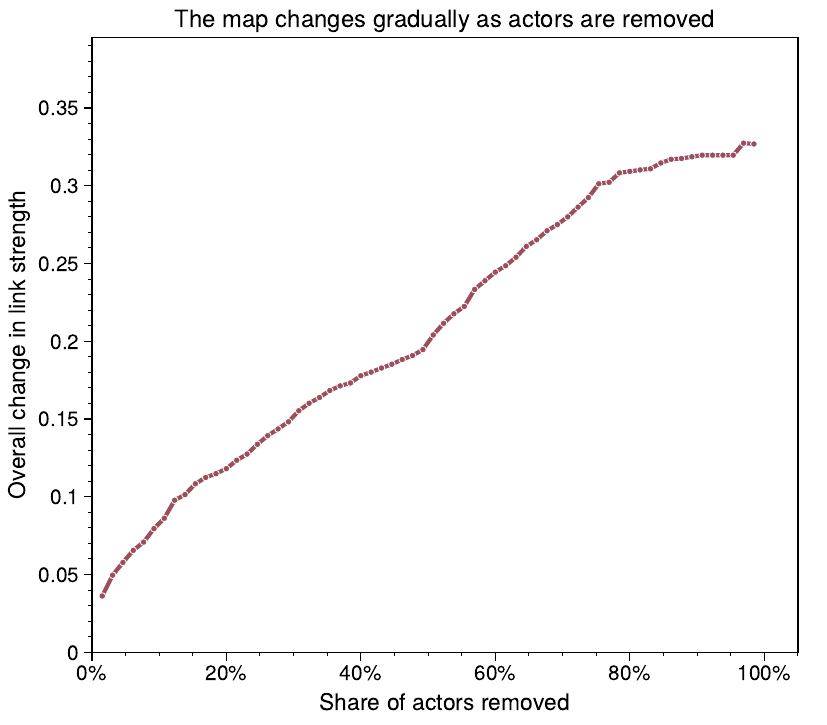}
 \caption{\textbf{No single participant determines the concern network.} Each step removes the participant whose exclusion maximizes root mean squared change from the original 66-participant proximity matrix, then rebuilds the network. Change accumulates gradually. We omit the undefined comparison after the final removal.}
 \label{SI-fig:sensitivity-analysis}
\end{figure}

In \cref{SI-fig:sensitivity-analysis}, the network changes gradually as we remove participants, indicating that no single participant drives the pattern. The comparison becomes less informative near the end of the sequence, when too few participants remain to estimate the original relations. This test shows that the full-record network draws on many participants; it does not establish that individual relation ranks or the network remain fixed over time.

\subsection{Aligning concerns with the subjects of adopted agreements}\label{SI-sec:outcome-positioning}
The inventory of adopted agreements contains all 584 records with titles from regular ATCMs 19--47 (1995--2025): 277 Measures, 135 Decisions, and 172 Resolutions. We begin in 1995 because Decision 1 (1995) established these three forms of agreement and the official register is complete for the period. Measures become legally binding after the required approvals. Decisions address internal organizational matters and are operative on adoption or another stated date. Resolutions are hortatory and not legally binding \cite{ATCMDecision1995Instruments}. We therefore group them only as \emph{adopted agreements}; the count does not imply equal legal effect. Agreements adopted at Special ATCMs, including the 1991 Madrid Protocol, are outside this inventory. Each record retains its form, number, year, and meeting because numbers repeat across years. The analysis reads these records directly from the archived 9 February 2026 official export and verifies its SHA-256 checksum, total counts, type counts, and counts by meeting.

The Secretariat's paper archive and official register use different classifications: 45 concerns for papers and 15 subjects for adopted agreements, which we call \emph{subjects}. Adopted agreements do not enter the network of concerns as additional concerns. We aggregate attention into the 15 subjects. This alignment compares the aligned paper totals with the unchanged official subject assignments of adopted agreements. It does not treat agreements as points in the network of concerns or use citations from agreements to papers. For a paper returned under several concerns, we sum its fractional weights within subjects; they still total one. Sixty agreements have more than one subject, giving 646 agreement--subject relations; an agreement with $k$ subjects contributes $1/k$ to each. The resulting weights also total one for every agreement.

Neither Secretariat source links the two classifications, so we construct a 45-to-15 alignment that assigns each concern to the subject that best matches its institutional function. This alignment changes how we aggregate attention; it does not recode adopted agreements. For example, Management Plans, Environmental Domains Analysis, Marine Protected Areas, and the general area-protection concern all enter ``Area protection and management''. \Cref{SI-tab:output-category-crosswalk} shows the complete alignment. \url{scripts/official_regular_atcm_outputs.py} freezes it, and \url{scripts/analyze_output_category_families.py} rebuilds the coarser network and agreement analysis.

We audited the concern-to-subject mapping using one AI agent (GPT-6.1 SOL, xhigh), with the author-coded assignments and prediction results withheld from it. The agent received only the 45 concern labels, the 15 subjects in randomized order, and an instruction to select each concern's principal institutional function. The agent matched 38 of 45 author-coded assignments (84.4\%). The agent reported high confidence for 28 assignments, medium confidence for 15, and low confidence for two. This audit measures agreement with the author-coded mapping and does not constitute human intercoder validation. The seven disagreements involved CEP Strategy Discussions, Marine Acoustics, Drilling, Operation of the Antarctic Treaty System Reports, Biological Prospecting, Inspections, and Operation of the CEP. We retain the author-coded mapping and treat these labels as boundary cases in the mapping-sensitivity analysis. \url{tools/build_crosswalk_blind_validation.py} creates the blinded packet, and \url{output/scientific_checks/crosswalk_blind_coder_a_comparison.csv} records the item-level audit.

We test dependence on this mapping using 27 boundary concerns and 50 plausible alternative assignments based on the concern labels. Each one-at-a-time test rebuilds concern-to-subject relations, historical networks, papers about the target subject, and papers weighted by linked subjects. We also evaluate 48 sparse joint mappings, in which each boundary concern has a 20\% reassignment probability, and 48 broad joint mappings drawn uniformly from the retained options. These ensembles generate stress scenarios rather than probabilities for the correct mapping. A post hoc failure search combines the individually most adverse assignments. Because it selects assignments using evaluation scores, it locates vulnerability rather than expected performance under coding error. The subjects of adopted agreements, evaluation meetings, model settings, and scoring remain fixed. \url{scripts/check_crosswalk_uncertainty.py} reproduces these checks.

\begin{table*}[!htbp]
\centering
\scriptsize
\caption{\textbf{An author-coded mapping assigns concerns to the 15 subjects of the official register.} The left column lists subjects from the Antarctic Treaty System official register. The right lists the concerns assigned to each. The mapping is unofficial.}
\label{SI-tab:output-category-crosswalk}
\begin{tabular}{@{}p{0.39\textwidth}p{0.53\textwidth}@{}}
\toprule
Subject & Included concerns \\
\midrule
Area protection and management & Area Protection and Management Plans General; Management Plans; Environmental Domains Analysis; Marine Protected Areas \\
Institutional \& legal matters & Operation of the ATS: General, Reports, and Secretariat; Cooperation with Other Organisations; Liability; Institutional and legal matters; Multiyear strategic workplan \\
Tourism and Non-Governmental Activities & Tourism and NG Activities; Site Guidelines for Visitors \\
Operational matters & Inspections; Search and Rescue; Operational issues; Safety and Operations in Antarctica; Emergency report and contingency planning \\
Environmental protection & Climate Change; Environmental Protection General; Environmental Monitoring and Reporting; CEP Strategy Discussions; Human Footprint and wilderness values; Operation of the CEP; State of the Antarctic Environment Report; Repair and remediation of environmental damage \\
Historic Sites and Monuments & Historic Sites and Monuments \\
Information exchange & Exchange of Information \\
Scientific cooperation & Science issues; Marine Acoustics; International Polar Year; Sub-glacial Lakes; Drilling; Biological Prospecting \\
Fauna and flora & Fauna and Flora General; Specially Protected Species; Nonnative Species and Quarantine \\
Marine pollution & Prevention of marine pollution \\
Environmental impact assessment & Environmental Impact Assessment: Other EIA Matters; Comprehensive Environmental Evaluations \\
General matters & Educational issues; Opening statements \\
Marine living resources & Marine living resources \\
Waste disposal and management & Waste management and disposal \\
Mineral resources & Mineral resources \\
\bottomrule
\end{tabular}
\end{table*}

\subsection{Retrospective predictions of agreement subjects}\label{SI-sec:prospective-output-forecast}
We predict the distribution of each meeting's adopted agreements across the 15 subjects, separately for Measures, Decisions, and Resolutions. The target is subject shares conditional on the observed number of agreements, not the number of agreements or the adoption of a particular proposal. For form of agreement $r$ at meeting $t$, let $q_{irt}$ be the observed share in subject $i$ and $\widehat{p}_{irt}$ its predicted share. The negative natural-log score is
\begin{equation}
\mathcal{L}_{rt}=-\sum_i q_{irt}\log\widehat{p}_{irt}.
\end{equation}
Lower scores are better. Reported changes subtract the reference model's score from the candidate's score within the same meeting. Thus, a negative change is an improvement, not a negative absolute loss.

For each prediction at ATCMs 29--47, a regularized Poisson model is fitted to ATCM 20 through the preceding meeting. The model includes subject indicators and numerical predictors standardized using the training records; fitted counts are normalized across subjects before scoring. The agreement-history reference uses the preceding five meetings of the same form of agreement in the target subject. These five-meeting histories, the one-meeting earlier-paper window, and regularization $0.1$ originate in the earlier combined-count specification's selection using all three forms of agreement at ATCMs 25--28. The distinction between Working and Information Papers and the log transformation were adopted later, after exploring ATCMs 29--47. We do not describe the revised specification as prospectively frozen or externally validated.

The target-subject model distinguishes fractionally counted Working Papers (WP) and Information Papers (IP), using each type's count in the target and immediately preceding meeting. It also includes the number of distinct participants across both types at each of those meetings. The archive's \texttt{party\_type} field supplies the WP/IP labels. Each retained paper contributes one total unit across its concerns, regardless of co-sponsorship, and the concern-to-subject mapping is unchanged. We apply $\log(1+x)$ to each attention count and participant-reach predictor before standardization, leaving the agreement-history count untransformed. The model can therefore estimate different relationships for the two documentary streams without assuming that document type perfectly distinguishes proposed action from reporting.

The full model adds target-meeting papers in every non-target subject with a positive link in the coarser network, up to 14 subjects. At each meeting, the network is rebuilt using only earlier papers. We set diagonal weights to zero, normalize each row's remaining proximity scores to one, and apply them to current paper counts. Attention to linked subjects still combines WP and IP counts; the log transformation is applied after their weighted sum. The earlier-information version omits all target-meeting attention and uses the corresponding preceding-meeting counts, participant reach, and attention to linked subjects, with the latter calculated from the network available before that source meeting. The paper files precede formal adoption, but concern assignments come from retrospective archive queries without assignment timestamps. Both versions therefore remain retrospective predictions, not reconstructed real-time forecasts.

We resample the 19 paired meeting-score differences 20,000 times for 95\% intervals and also report circular moving-block intervals with blocks of four meetings. Contrasts among forms of agreement resample the paired differences between their gains. Intervals do not account for choosing a specification after examining this evaluation period or for all alternatives explored. We fit the same document-type specification to all participants, time-varying Consultative Parties, and the fixed original-signatory set, rebuilding attention and the network for each participant set while keeping all adopted agreements as targets. \url{scripts/analyze_document_type_prediction.py} produces these fits and their contrasts.

We additionally compare the fitted model with simple historical subject frequencies. These use either all earlier training meetings or the preceding five meetings, add $0.5$ units of agreement weight to each subject to avoid zero probabilities, and normalize to one. This reference has no documentary predictors. \url{scripts/document_type_prediction_tables.py} produces the reference comparisons and manuscript tables. A separate exploratory comparison selects among document representations using only prediction scores from earlier meetings; because the candidate representations were developed after inspecting the evaluation period, this is not external validation (\cref{SI-sec:document-type-prediction}).

Earlier combined-count checks are retained below as specification history. They vary windows, regularization, linked-subject definitions, concern-to-subject mappings, participant sets, and information timing, and include 200 coherent row-and-column label shuffles of the concern network. They are not automatically robustness checks of the revised document-type model. The original scripts in \url{scripts/analyze_resolution_attention_forecast.py}, \url{scripts/evaluate_nested_attention_forecast.py}, \url{scripts/adversarial_forecast_checks.py}, and \url{scripts/audit_snapshot_nowcast_sensitivity.py} reproduce those earlier comparisons.

\subsection{Auditing formal citations among Measures}\label{SI-sec:formal-inheritance-audit}
The prediction asks whether attention helps locate adopted agreements. A separate audit asks what formal record later Measures cite. We froze the adopted body text from the official detail page of every one of the 277 Measures on 16 August 2026. The pinned JSON Lines (JSONL) corpus records each official URL, retrieval date, extraction status, and body hash; an adjacent manifest records the corpus SHA-256 and verifies the source inventory and all 277 records. The audit consumes this corpus without live requests. We extract exact references of the form \emph{Measure}, \emph{Decision}, or \emph{Resolution} followed by a number and year. Agreement numbers restart each year, so type, number, and year jointly identify every source. A reference counts only when it resolves to an earlier entry in the pinned official inventory. Repeated mentions of the same source within one Measure count as one link.

The audit reports the number of later Measures citing each source type, the number of distinct earlier agreements cited, and the number of source--Measure links. A later Measure can cite several source types, so the percentages in \cref{SI-fig:measure-formal-citations} overlap. These exact references provide a conservative record of formal citation. They do not establish that the earlier agreement caused the later one, and a policy sequence can omit an intermediate source from its final adopted body text. We also screen Measure titles for ``management plan'' and ``revised management plan'' to describe the recurring content of the Measure record. \url{scripts/plot_formal_lineage_overview.py} reproduces the lineage overview.

To examine whether heavily discussed Resolutions later receive a citation from a Measure, we retain Resolutions with ten subsequent regular meetings in the archive. A Resolution enters the linked set only when a Measure adopted within those ten meetings contains an exact year-qualified citation to it. For each linked Resolution, we compare same-subject attention with uncited Resolutions sharing at least one subject and report its percentile within that set. We count time in meetings rather than calendar years. The linked sample is small, so this is a descriptive boundary test rather than a fitted transition model. \url{scripts/audit_resolution_measure_ladders.py} and \url{scripts/analyze_resolution_hardening_attention.py} reproduce the audit.

\subsection{Paper--agreement associations}\label{SI-sec:attention-output-models}
The subject--meeting analysis orders the record by ATCM number, so one step is the immediately preceding meeting even when more than one calendar year elapsed. For subject $i$ at ATCM $t$, we measure three earlier histories: fractionally counted papers in $i$; papers in its five most closely linked subjects, weighted using the network available before each source meeting; and adopted agreements in $i$. For the primary one-meeting analysis, each history comes from ATCM $t-1$, and its linked-subject weights use the network built through $t-2$. We add one and take the logarithm.

A Poisson pseudo-maximum-likelihood model uses subject and meeting fixed effects. The reported rate ratios give the change associated with doubling one plus an earlier count. We cluster covariance by subject and meeting. Because there are only 15 subject clusters, the intervals require caution. They also condition on the author-coded concern-to-subject mapping and estimated network; the models do not propagate uncertainty in those objects.

We repeat the analysis after combining the preceding one, two, three, five, eight, or ten meetings. The available meeting set therefore shrinks as the window grows. To separate window length from sample period, we repeat every horizon on the 19 meetings shared by the ten-meeting model (ATCMs 29--47). Composition checks exclude General matters, Area protection and management, or both. The latter matters because recurring area and management-plan work makes up almost half of the subject weight of adopted agreements. All comparisons describe temporal association between attention and adopted agreements, not a causal effect of papers on agreements. \url{scripts/analyze_output_category_families.py} reproduces the models.

\section{Supplementary Results}\label{SI-sec:si-results}
\subsection{Consultative and Non-Consultative Parties occupy different regions}\label{SI-sec:group-positioning-check}
\begin{figure*}[!htbp]
\centering
\includegraphics[width=\textwidth]{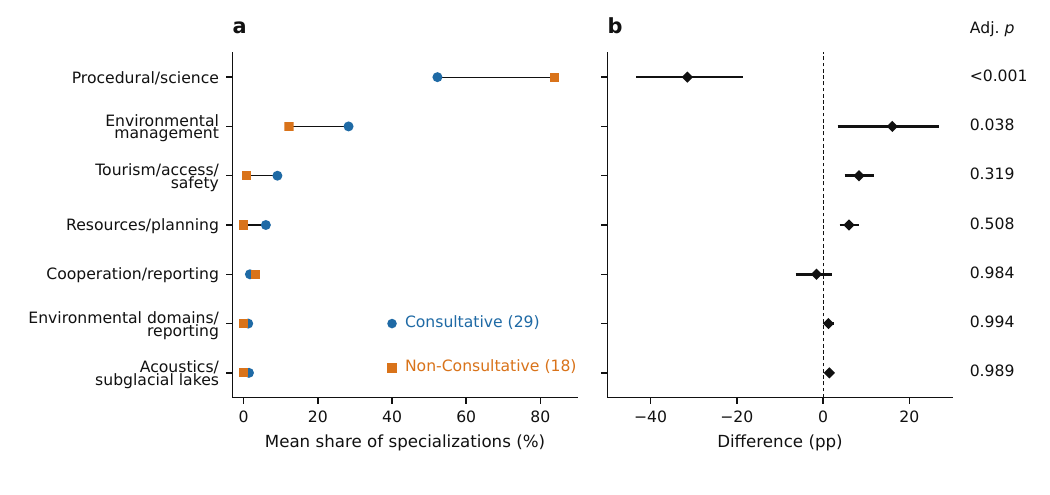}
\caption{\textbf{State portfolios concentrate in different regions.} (a) Mean regional shares of each state's specializations, for 29 Consultative and 18 Non-Consultative Parties. (b) Consultative minus Non-Consultative shares in percentage points, with unadjusted descriptive 95\% ranges from 10,000 within-group state resamples. Regional $p$ values use the largest absolute difference across the seven regions in each of 99,999 label permutations.}
\label{SI-fig:group-positioning}
\end{figure*}

To compare the composition of state portfolios, we express each state's portfolio as shares across the seven displayed regions (\cref{SI-sec:cpm-reading}). Each of the 29 Consultative and 18 Non-Consultative Parties receives equal weight, using the full-record classification in \cref{SI-sec:group-specialization}. The Euclidean distance between the two mean compositions is $0.3687$. Four of 99,999 label permutations give an equal or larger distance (plus-one Monte Carlo $p=0.00005$). Procedural and scientific concerns account for 52.2\% of the mean Consultative portfolio and 83.7\% of the Non-Consultative portfolio; environmental management accounts for 28.3\% and 12.2\% (\cref{SI-fig:group-positioning}). Comparing shares across all 45 concerns also gives $p=0.00001$, the simulation's minimum value.

Median submission counts are 148 and 10.5, and median first recorded submission years are 1981 and 2004, respectively. We also permute labels within joint strata of submission-volume tertile and first-submission era (through 1990, 1991--2010, after 2010). The mixed-group strata contain 13 states and permit 30 distinct assignments; exact enumeration gives $p=1/30=0.033$.

\subsection{Bridge ranks on the displayed and rebuilt networks}\label{SI-sec:bridge-stability}
In each of 1,000 resamples of the 66 complete participant records, we recompute RPA and proximity using the definitions in \cref{SI-sec:map-construction}. We calculate betweenness ranks on the displayed links with their new positive weights and on a network rebuilt from the same resample. Rebuilding forms a new maximum spanning tree and reselects supplementary links from 500 inner participant resamples, retaining links with a positive 2.5th-percentile proximity and endpoint-top-five frequency of at least 0.60. Path lengths are $-\log\phi$, with unit proximity clipped to $1-10^{-9}$ to keep lengths positive. Ranks cover all present concerns; ties receive average ranks.

\begin{figure*}[!htbp]
\centering
\includegraphics[width=\textwidth]{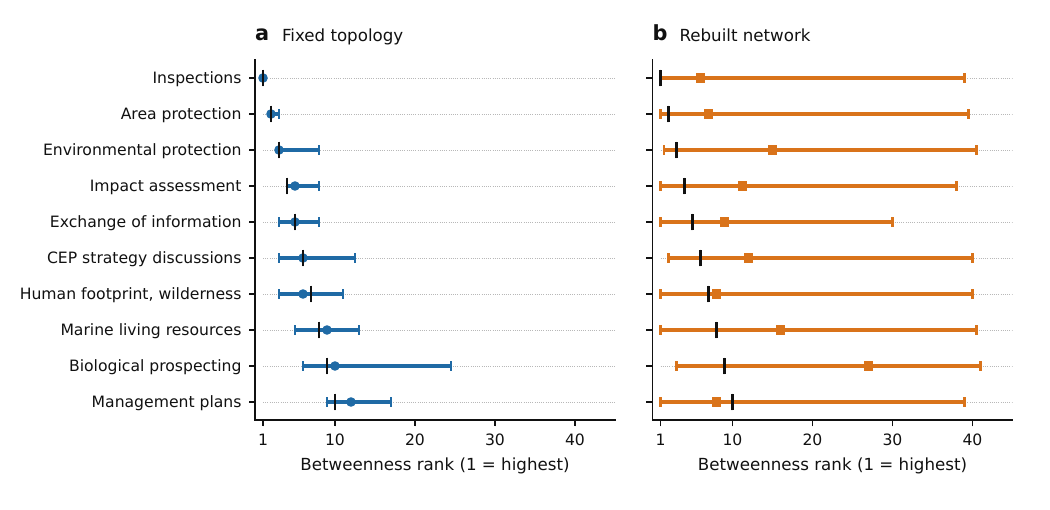}
\caption{\textbf{Bridge ranks vary more when links are rebuilt.} The ten concerns with highest observed betweenness appear in the same order and on the same rank scale. (a) Weights recomputed on the displayed links. (b) Links rebuilt in each participant resample. Symbols mark median ranks, bars central 95\% ranges across 1,000 resamples, and black ticks observed ranks. Rank 1 is highest; ties receive average ranks.}
\label{SI-fig:bridge-stability}
\end{figure*}

Inspections and area protection rank first and second in the displayed network. With its links fixed, their central 95\% rank ranges are 1--1 and 2--3. Rebuilding gives median ranks of 6 and 7, with ranges 1--39 and 1--39.5 (\cref{SI-fig:bridge-stability}). They appear among the four highest-ranked concerns in 42.4\% and 38.2\% of rebuilt draws, and both are present in every draw.

\subsection{Tourism, marine living resources, and climate change}\label{SI-sec:concern-case-profiles}
\Cref{SI-tab:concern-case-profiles} places three concerns within the network and compares their specialists, distances to earlier portfolios, and adopted outputs. For each concern, we use its distance to the prior five-meeting portfolio (\cref{SI-sec:hazard-model}), with networks built from records through that prior window. We include a participant-window when the concern has already appeared in the archive and the participant has a nonempty portfolio without that specialization. Distances are averaged within participant before participants receive equal weight. Climate change lies farther from earlier portfolios on average than tourism or marine living resources.

Agreement counts use the official tourism and marine-living-resources subjects at regular ATCMs in 1995--2025 (\cref{SI-sec:outcome-positioning}). Climate change has no dedicated agreement subject; we count titles containing ``climat'', ``warming'', or ``carbon''. This is an incomplete climate subset. Marine counts cover ATCM instruments and exclude CCAMLR conservation measures.

\begin{table*}[!htbp]
\centering
\small
\caption{\textbf{Three concerns connect different portfolios and forms of agreement.} Papers and specialists use the full 1961--2025 record. CP/NCP/Org gives specialist counts among 29 Consultative Parties, 18 Non-Consultative Parties, and 19 organizations and institutional bodies. Distance is the actor-weighted historical mean of $1-\max\phi$. M/R counts Measures and Resolutions adopted at regular ATCMs in 1995--2025.}
\label{SI-tab:concern-case-profiles}
\begin{tabular}{@{}llrccc@{}}
\hline
Concern & Region & Papers & CP/NCP/Org & Distance & M/R \\
\hline
Tourism & Tourism/access/safety & 684 & 8/0/3 & 0.685 & 2/42 \\
Marine living resources & Resources/planning & 48 & 8/0/1 & 0.686 & 0/5 \\
Climate change & Environmental management & 207 & 7/2/5 & 0.752 & 0/6$^{\dagger}$ \\
\hline
\end{tabular}
\par\smallskip{\footnotesize $^{\dagger}$ Climate-explicit titles; incomplete subset.}
\end{table*}

\subsection{Prediction performance differs among forms of agreement}\label{SI-sec:forecast-results}
The separate-document-type specification improves Resolution and Measure predictions relative to the original agreement-history model (\cref{SI-tab:output-forecast-scores}). Target-subject information supplies most of these gains; the intervals for the additional contribution of linked subjects span zero for all three forms of agreement. The Resolution gain exceeds the Decision gain by $0.122$ (95\% interval for the difference $[-0.203,-0.042]$). The Resolution--Measure difference is less certain ($-0.050$, $[-0.122,0.023]$). We therefore do not interpret a significant Resolution gain as evidence that attention informs only soft law.

\begin{table*}[!htbp]
\centering\scriptsize
\caption{\textbf{Separating Working and Information Papers improves Resolution and Measure category prediction relative to the original output-history model.} Mean changes in negative natural-log score over ATCMs 29--47, with 95\% paired meeting-bootstrap intervals. Negative values favour the added predictors. Target-category counts distinguish Working and Information Papers and are log-transformed; nearby attention combines document types. This specification was adopted after retrospective exploration, and the intervals do not adjust for model selection.}
\label{SI-tab:output-forecast-scores}
\begin{tabular}{@{}lccc@{}}
\toprule
Form of output & Target papers vs history & Target + nearby vs history & Nearby contribution \\
\midrule
Measure & $-0.054\;[-0.088,-0.023]$ & $-0.056\;[-0.090,-0.025]$ & $-0.002\;[-0.004,0.001]$ \\
Decision & $0.015\;[-0.034,0.063]$ & $0.016\;[-0.032,0.065]$ & $0.001\;[-0.003,0.005]$ \\
Resolution & $-0.095\;[-0.150,-0.042]$ & $-0.106\;[-0.166,-0.046]$ & $-0.011\;[-0.027,0.008]$ \\
\bottomrule
\end{tabular}
\end{table*}

\begin{figure*}[!htbp]
\centering
\includegraphics[width=\textwidth]{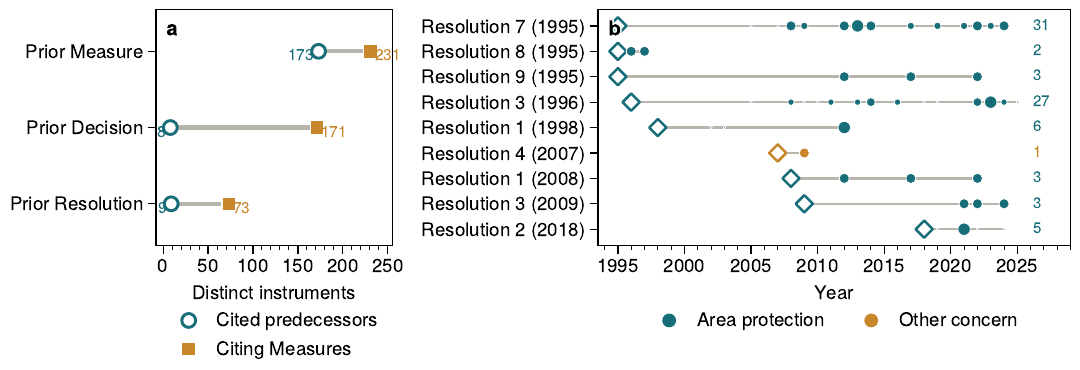}
\caption{\textbf{Exact citations link later Antarctic Treaty System Measures to earlier Measures, Decisions, and Resolutions.} \textbf{a}, 231 distinct later Measures cite 173 distinct earlier Measures, 171 distinct later Measures cite at least one of eight Decisions, and 73 distinct later Measures cite at least one of nine Resolutions. Rows overlap because a Measure can cite several source types. \textbf{b}, each row traces one cited Resolution. The diamond marks its adoption year, and circles mark later years when Measures cite it; circle area scales with the number of citing Measures in that year. The number at right gives the number of distinct Measures that cite that Resolution. Eight of the nine source Resolutions carry the official Area protection and management subject; teal marks those eight lineages, and orange marks the other lineage. Across all nine lineages, 71 of the 73 distinct citing Measures also carry that subject. Exact citations establish formal citation links. They do not show that a cited agreement caused a later Measure or that any Measure entered into force.}
\label{SI-fig:formal-lineage-overview}
\end{figure*}

Later Measures cite nine source Resolutions. Across these sources, the first Measure to cite each Resolution followed it by a median of four Antarctic Treaty Consultative Meetings (range, 1 to 16 meetings; $n = 9$). Later Measures cited these Resolutions over decades (\cref{SI-fig:formal-lineage-overview}b). Eight of the nine source Resolutions and 71 of the 73 citing Measures carry the official Area protection and management subject. Each source Resolution and the later Measures that cite it form a formal citation lineage. These nine lineages document repeated formal citations concentrated in one policy area, with three citation links outside it; they do not establish a general mechanism in which Resolutions lead to Measures.

The three forms of agreement pose different prediction problems. The effective number of occupied subjects, calculated as the exponential of Shannon entropy, is 10.55 for Resolutions, 3.14 for Decisions, and 1.38 for Measures. Area protection and management accounts for $93.1\%$ of fractional Measure weight, while Institutional and legal matters accounts for $73.0\%$ of Decision weight. The largest Resolution subject, Tourism and Non-Governmental Activities, accounts for only $22.1\%$. Earlier agreements are therefore already a strong guide to the narrow Measure and Decision distributions.

Under the earlier combined-count specification, the Resolution score change is $-0.073$ in the ten meetings at or below median subject entropy ($[-0.141,-0.006]$) and $-0.055$ in the nine meetings above it ($[-0.125,0.021]$). Their difference is $0.018$ ($[-0.080,0.119]$). The correlation between score change and entropy is $-0.043$ ($[-0.417,0.318]$). These earlier estimates describe the combined-count model; the same concentration check has not been rerun for the document-type specification.

Antarctic Specially Protected Areas (ASPAs) are designated under Annex V of the Environmental Protocol to protect outstanding environmental, scientific, historic, aesthetic, or wilderness values, or ongoing or planned scientific research \cite{ATSProtectedAreas}.

\begin{figure}[!htbp]
\centering
\includegraphics[width=0.82\columnwidth]{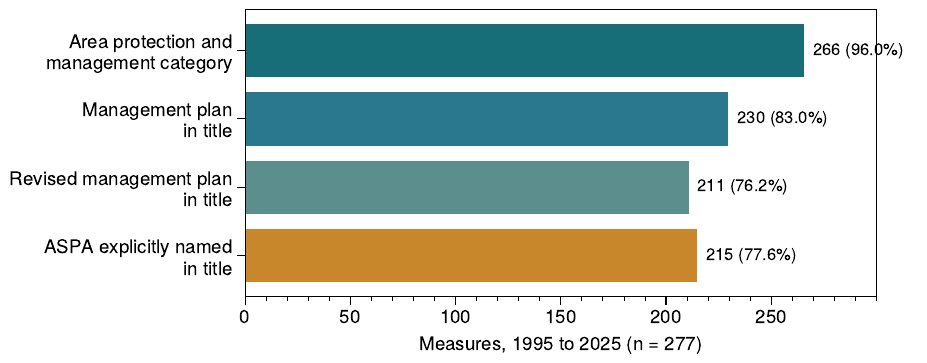}
\caption{\textbf{Measures concentrate in recurring area-protection and management-plan work.} Bars show overlapping indicators among 277 Measures adopted at regular ATCMs from 1995 to 2025. Area protection and management is a subject of the official register; the remaining indicators come from Measure titles. The ASPA bar combines 160 titles that spell out ``Antarctic Specially Protected Area'' and 55 that use the acronym.}
\label{SI-fig:measure-composition}
\end{figure}

In the earlier combined-count analysis of Measures, the agreement-history model based on the preceding five meetings gives a mean prediction score of $0.388$. Models using prior Decisions and Resolutions alone score $1.134$, all earlier agreements score $0.486$, and papers about the target and linked subjects without Measure history score $0.759$. Adding Decisions and Resolutions to the Measure agreement-history model worsens the score by $0.011$ ($[0.000,0.023]$). Adding papers to that model changes it by $-0.001$ ($[-0.023,0.020]$). The recurrence is consistent with the content of the record: 266 of 277 Measures carry the Area protection and management subject, 230 titles concern a management plan, 211 concern a revised management plan, and 215 explicitly identify an ASPA (\cref{SI-fig:measure-composition}).

In the earlier combined-count specification, the Resolution result is not reproduced by generic activity outside the target subject. Adding all other papers to the target-subject paper model changes the score by $0.001$ ($[-0.001,0.003]$), compared with $-0.023$ for papers weighted by closeness to the target. Omitting each subject from the score leaves the mean gain from linked subjects negative in all 15 cases, ranging from $-0.033$ to $-0.016$. Every one of the 16 combinations of earlier-agreement and paper windows gives a negative mean gain, as do all five tested numbers of linked subjects and all four regularization values. All six projections through the original 45-concern network improve the mean score. Moving-block bootstraps of one to four meetings preserve the negative mean, although their intervals cross zero. The gain is $-0.008$ in ATCMs 29--38 ($[-0.038,0.025]$) and $-0.040$ in ATCMs 39--47 ($[-0.083,-0.005]$). When every setting is instead reselected within each training period using earlier meetings, the combined Resolution change remains similar at $-0.067$, but its interval crosses zero ($[-0.131,0.004]$). These checks apply to the earlier combined-count specification. They do not establish robustness of the revised document-type model; in particular, the reselection result is not a selection test of the revised feature representation.

\subsection{Separating documentary streams improves prediction}\label{SI-sec:document-type-prediction}
The revised specification follows exploratory comparisons of short-gap persistence, transformations of paper counts, relative attention, alternative fitting objectives, regularization, recent training windows, and document metadata. These comparisons reused ATCMs 29--47. We selected the separate-WP/IP representation because it improved prediction and retained a substantive distinction recorded by the archive, not because it had been specified before examining the evaluation period. The same representation and settings were then applied to all three forms of agreement and both restricted participant sets. Earlier specifications and their results remain available rather than being replaced in the analysis archive.

The fitted coefficients distinguish the documentary streams (\cref{SI-tab:document-type-coefficients}). Working Paper coefficients for Measures are positive and Information Paper coefficients negative in all 19 expanding-window fits, under both timing specifications. With preceding-meeting papers alone, their median standardized coefficients are $0.832$ and $-0.106$, respectively. Both streams receive positive coefficients for Resolutions in this specification ($0.167$ for Working Papers and $0.247$ for Information Papers). In the target- and preceding-meeting specification, target-meeting coefficients for both types and the preceding-meeting Information Paper coefficient remain positive; the preceding-meeting Working Paper coefficient changes sign across fitted windows. Decision coefficients are positive for both streams, although separating the types does not improve their prediction score. These are conditional predictive weights. \url{scripts/report_document_type_coefficients.py} exports the coefficients and verifies that the fits reproduce the archived predictions.

\begin{table*}[!htbp]
\centering
\caption{\textbf{Working Papers and Information Papers carry different predictive weights.} Entries are median standardized log-link coefficients, followed by the minimum and maximum across the 19 expanding-window fits used to predict ATCMs 29--47. Paper counts are transformed as $\log(1+x)$ and standardized within each training set. Coefficients are conditional on the other predictors, including agreement history, participant reach, attention to connected subjects, and subject indicators. Ranges describe variation across fitted windows, not confidence intervals.}
\label{SI-tab:document-type-coefficients}
\small
\begin{tabular}{lrrr}
\hline
Predictor & Measures & Decisions & Resolutions \\
\hline
\multicolumn{4}{l}{\emph{Target- and preceding-meeting papers}} \\
Target-meeting WP & $0.764\,[0.573,0.868]$ & $0.275\,[0.184,0.364]$ & $0.241\,[0.211,0.300]$ \\
Target-meeting IP & $-0.217\,[-0.256,-0.105]$ & $0.076\,[0.027,0.117]$ & $0.122\,[0.041,0.135]$ \\
Preceding-meeting WP & $0.532\,[0.193,0.568]$ & $0.178\,[0.157,0.281]$ & $-0.003\,[-0.122,0.107]$ \\
Preceding-meeting IP & $-0.110\,[-0.206,-0.035]$ & $0.081\,[0.003,0.118]$ & $0.172\,[0.137,0.206]$ \\
\hline
\multicolumn{4}{l}{\emph{Preceding-meeting papers only}} \\
Preceding-meeting WP & $0.832\,[0.376,0.887]$ & $0.301\,[0.274,0.359]$ & $0.167\,[0.023,0.245]$ \\
Preceding-meeting IP & $-0.106\,[-0.201,-0.020]$ & $0.133\,[0.062,0.169]$ & $0.247\,[0.215,0.278]$ \\
\hline
\end{tabular}
\end{table*}

\begin{figure*}[!htbp]
\centering
\includegraphics[width=\textwidth]{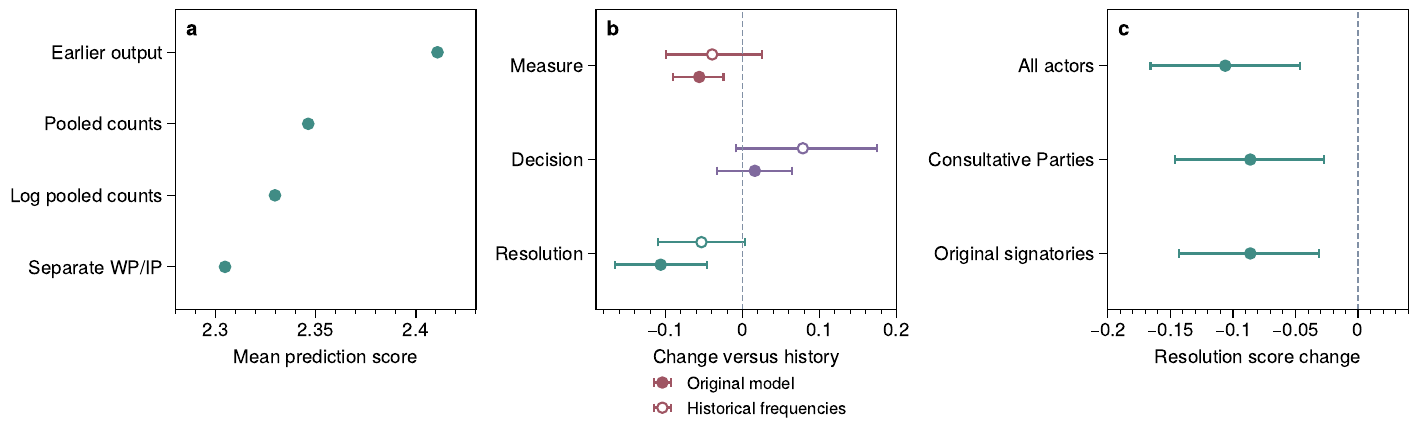}
\caption{\textbf{Document type improves retrospective performance, while the historical reference bounds the gain.} (a) Mean Resolution prediction scores for the original history model, combined raw paper counts, combined log-transformed counts, and separate log-transformed Working Paper and Information Paper counts. (b) Same-meeting gains from the revised model relative to either the original fitted history model or smoothed subject frequencies from all earlier training agreements. (c) Resolution gains relative to the original history model after rebuilding documentary predictors for all participants, time-varying Consultative Parties, or original signatories. Bars in (b) and (c) are paired meeting-bootstrap 95\% intervals; they do not adjust for model selection. All comparisons evaluate ATCMs 29--47.}
\label{SI-fig:document-type-prediction}
\end{figure*}

\begin{figure}[!htbp]
\centering
\includegraphics[width=\columnwidth]{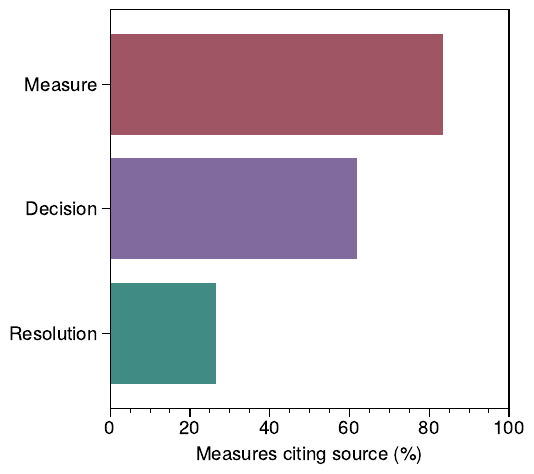}
\caption{\textbf{Measures frequently cite earlier adopted agreements.} Proportion of 277 Measures from regular ATCMs, 1995--2025, citing earlier Measures, Decisions, or Resolutions in their adopted body text. References must identify an agreement by form, number, and year and match the official inventory. A Measure can cite several source types, so the percentages overlap.}
\label{SI-fig:measure-formal-citations}
\end{figure}

For Resolutions, the original combined-count model has a mean log score of $2.346$, log-transformed combined counts score $2.330$, and separate log-transformed WP/IP counts score $2.305$ (\cref{SI-fig:document-type-prediction}a). Separating the types improves the score relative to log-transformed combined counts by $0.025$ ($[-0.042,-0.008]$), improving 14 of 19 meetings. With preceding-meeting information only, the additional change is $-0.022$ ($[-0.034,-0.012]$), improving 17 of 19 meetings. The four-meeting-block intervals remain below zero for both comparisons. This result does not require a semantic classifier: titles with explicit action words gave no clear improvement, and WP/IP labels are taken directly from archive metadata. Nor does it imply that every Working Paper seeks adoption or every Information Paper merely reports existing work.

The original agreement-history reference scores $2.411$ for Resolutions, giving a revised same-meeting gain of $-0.106$ ($[-0.166,-0.046]$) and a preceding-meeting gain of $-0.092$ ($[-0.141,-0.040]$), each improving 15 of 19 meetings. Against a smoothed distribution of subjects in all earlier training agreements, however, the changes are $-0.053$ ($[-0.110,0.003]$) and $-0.040$ ($[-0.096,0.015]$), respectively (\cref{SI-tab:document-type-references}). For the same-meeting comparison, the circular four-meeting-block interval is $[-0.102,-0.006]$, whereas the independent-meeting interval crosses zero; the uncertainty conclusion therefore depends on the resampling scheme. Better absolute performance does not imply an equally large or equally certain contribution beyond every representation of institutional history.

Measures also improve in the revised same-meeting specification relative to the original agreement-history model ($-0.056$, $[-0.090,-0.025]$). Their preceding-meeting change is smaller ($-0.023$, $[-0.046,-0.001]$), and its four-meeting-block interval crosses zero ($[-0.057,0.006]$). Against the all-earlier-frequency reference, both Measure intervals span zero (\cref{SI-fig:document-type-prediction}b). These results rule out carrying the earlier specification's clean Resolution--Measure separation into the revised analysis. Decisions show no clear improvement over either historical reference.

A selection check among the document representations chooses log-transformed combined counts using only ATCMs 25--28. Reselecting before each later prediction using only earlier prediction scores selects separate WP/IP counts in 14 of 19 meetings, but the resulting mean change relative to the log-combined model is uncertain ($-0.016$, $[-0.033,0.003]$). All reported intervals are unadjusted for the broader exploration. The evidence supports a promising retrospective representation of documentary attention, not an independently established improvement in future prediction. \url{scripts/explore_document_role_predictions.py} records the representation and selection comparisons; the revised estimates, participant restrictions, and figures are reproduced by \url{scripts/analyze_document_type_prediction.py}, \url{scripts/document_type_prediction_tables.py}, and \url{scripts/plot_document_type_prediction.py}.

\begin{table*}[!htbp]
\centering\scriptsize
\caption{\textbf{The gain depends on the historical comparison and information cutoff.} Entries compare the separate-document-type model with either the original regularized output-history model or a smoothed historical frequency distribution, using all earlier training outputs and half an instrument of category weight added to each category. Intervals resample paired meetings. The frequency reference uses no paper information.}
\label{SI-tab:document-type-references}
\begin{tabular}{@{}llcc@{}}
\toprule
Output & Information cutoff & Original history model & Historical frequencies \\
\midrule
Measure & Same meeting & $-0.056\;[-0.090,-0.025]$ & $-0.039\;[-0.100,0.025]$ \\
Measure & Previous meeting & $-0.023\;[-0.046,-0.001]$ & $-0.006\;[-0.076,0.072]$ \\
Decision & Same meeting & $0.016\;[-0.032,0.065]$ & $0.078\;[-0.009,0.175]$ \\
Decision & Previous meeting & $0.014\;[-0.029,0.057]$ & $0.077\;[-0.011,0.174]$ \\
Resolution & Same meeting & $-0.106\;[-0.166,-0.046]$ & $-0.053\;[-0.110,0.003]$ \\
Resolution & Previous meeting & $-0.092\;[-0.141,-0.040]$ & $-0.040\;[-0.096,0.015]$ \\
\bottomrule
\end{tabular}
\end{table*}

\subsection{Gap checks on the earlier combined-count model}\label{SI-sec:attention-gap-sensitivity}

Using the earlier combined-count specification, we examine whether a subject's attention remains informative when no new paper appears for one or two meetings. This exploratory comparison retains all participants, subject indicators, five meetings of Resolution history, participant reach, and attention to linked subjects, with regularization fixed at $0.1$. We add either an indicator that remains active through one or two consecutive zero-paper meetings, the last positive paper count carried forward through those gaps, or a sum of paper counts that declines exponentially with a half-life of one or two meetings. Indicator models also include the number of meetings since the last positive paper count, capped at five. As a comparison, we replace the preceding-meeting paper count with sums over two, three, or five preceding meetings, leaving participant reach unchanged. Each model is fitted on earlier meetings and evaluated at ATCMs 29--47. We repeat the comparison without target-meeting information, using the preceding meeting's attention and lagged versions of the added predictors. Gap lengths refer to meetings, not calendar years; these are subject-level gaps, not interruptions in individual participants' specialization.

The reconstructed combined-count model reproduces the earlier Resolution gain relative to agreement history ($-0.0646$, 95\% meeting-bootstrap interval $[-0.1142,-0.0144]$, improving 15 of 19 meetings). Relative to that full attention model, none of the gap alternatives shows a clear additional gain (\cref{SI-tab:attention-gap-sensitivity}). The largest observed improvement is $-0.0041$ for same-meeting predictions and $-0.0068$ when only earlier attention is used; both intervals span zero. These are the best observed estimates among several alternatives, not independently selected improvements. Selecting among the alternatives and the unchanged model using Resolution scores at ATCMs 25--28 instead chooses the five-meeting paper window for same-meeting prediction (later score change $0.0015$, $[-0.0071,0.0108]$), and two-meeting exponential decay for earlier-information prediction ($-0.0056$, $[-0.0138,0.0038]$). Neither provides a clear gain. Ten of the 15 subjects have papers at every evaluation meeting, limiting the opportunities for gap retention to change the predictors. These gap results concern the earlier combined-count model, not the revised separate-document-type specification; they reuse the evaluation period and do not provide external validation. \url{scripts/explore_full_attention_gaps.py} reproduces the analysis.

\begin{table*}[!htbp]
\centering
\scriptsize
\caption{\textbf{Short-gap persistence does not clearly improve the earlier combined-count Resolution model.} Entries give mean changes in negative natural-log prediction score, with 95\% intervals from 20,000 paired meeting-bootstrap resamples over ATCMs 29--47. Negative differences favour the alternative over the full attention model with the corresponding information cutoff, not over agreement history alone. All variants are reported, including those that worsen the mean score.}
\label{SI-tab:attention-gap-sensitivity}
\begin{tabular}{lcc}
\toprule
Attention alternative & Same-meeting information & Earlier information only \\
\midrule
Active through one silent meeting & $-0.0041\;[-0.0139,0.0065]$ & $-0.0057\;[-0.0146,0.0048]$ \\
Active through two silent meetings & $-0.0036\;[-0.0126,0.0062]$ & $-0.0068\;[-0.0160,0.0050]$ \\
Carry count through one silent meeting & $0.0011\;[0.0001,0.0022]$ & $-0.0007\;[-0.0028,0.0015]$ \\
Carry count through two silent meetings & $0.0011\;[0.0001,0.0022]$ & $-0.0007\;[-0.0028,0.0015]$ \\
Exponential decay, one-meeting half-life & $-0.0008\;[-0.0030,0.0013]$ & $-0.0043\;[-0.0099,0.0024]$ \\
Exponential decay, two-meeting half-life & $-0.0023\;[-0.0069,0.0023]$ & $-0.0056\;[-0.0138,0.0038]$ \\
Two-meeting paper window & $0.0001\;[-0.0057,0.0063]$ & $0.0013\;[-0.0077,0.0112]$ \\
Three-meeting paper window & $-0.0015\;[-0.0100,0.0081]$ & $-0.0014\;[-0.0132,0.0128]$ \\
Five-meeting paper window & $0.0015\;[-0.0071,0.0108]$ & $0.0015\;[-0.0112,0.0157]$ \\
\bottomrule
\end{tabular}
\end{table*}

\subsection{Earlier timing and mapping checks}\label{SI-sec:forecast-audits}
The prediction checks in this subsection use the earlier combined-count specification. Their numerical results are retained for transparency and do not establish timing or mapping robustness of the revised document-type model. The agreement-to-attention analysis at the end is a separate association test, unchanged by the prediction revision.
In the earlier combined-count specification, papers from the preceding meeting improve only the Resolution prediction. Papers about the target subject change the Resolution score by $-0.042$ ($[-0.076,-0.006]$), and papers about the target and linked subjects change it by $-0.082$ ($[-0.128,-0.033]$). The combined model improves 15 of 19 meetings, and moving-block intervals of one to four meetings remain below zero. The exact two-sided sign-flip probability is $0.005$. For Measures and Decisions, the corresponding preceding-meeting changes are $0.007$ ($[-0.022,0.043]$) and $0.017$ ($[-0.025,0.058]$). The result therefore retains the separation among forms of agreement.

Future attention provides a weaker persistence check. Attention from the next meeting changes the Resolution score by $-0.055$ across 18 meetings and improves 12. Moving-block intervals for blocks of one and two meetings cross zero, and the exact two-sided sign-flip probability is $0.115$. This result does not separate a stable agenda from directional prediction, but the clearer preceding-meeting association shows that focal-meeting attention is not necessary for the main pattern.

The January 2024 snapshot preserves one concern for 6,230 papers that also appear in the 2026 reconstruction. Every retained concern matches its later record, including all 3,644 matched papers from ATCMs 29--45. Over these 17 evaluation meetings, the older labels preserve the target-paper and combined Resolution gains. The improvement from linked subjects changes from $-0.018$ ($[-0.043,0.006]$) under the multi-label reconstruction to $0.003$ ($[-0.002,0.008]$) under the single-label snapshot. The snapshot cannot date its labels and excludes ATCMs 46--47. It supports stability of the overall attention association but weakens attribution to attention to linked subjects alone.

Changes to the concern-to-subject mapping identify a separate boundary. All 50 one-at-a-time reassignments retain a negative estimate for papers about the target and linked subjects, ranging from $-0.075$ to $-0.049$, and 48 intervals remain below zero. All 48 sparse joint mappings retain negative estimates from $-0.085$ to $-0.029$, with 34 intervals below zero. All 48 broad joint mappings also retain negative estimates from $-0.060$ to $-0.008$, with 15 intervals below zero. The improvement from linked subjects remains negative in every isolated and sampled mapping, although its interval often crosses zero. A post hoc stack of 19 individually adverse assignments moves the combined estimate to $0.009$ ($[-0.033,0.053]$) and the improvement from linked subjects to $0.001$ ($[-0.004,0.006]$). Local and sampled mapping uncertainty does not explain the main point estimate, but a coordinated alternative mapping can remove it. The separate AI-agent audit agrees with 38 of 45 author-coded assignments (84.4\%), with seven disagreements among the boundary concerns examined here.

\paragraph{Agreement-to-attention feedback.} We test the reverse direction on the same 15-subject by ATCM-meeting panel used for the attention-to-Resolution forecast, with fractional paper attention as the outcome. For each subject and meeting, a Poisson pseudo-maximum-likelihood (PPML) model regresses paper attention summed over the next five meetings on current subject-weighted counts of Measures, Decisions, and Resolutions, controlling for the subject's own preceding five-meeting attention total, with subject and origin-meeting fixed effects and covariance clustered two ways by subject and meeting. Subject-weighted counts of Resolutions have a positive association with later attention (rate ratio per doubling $1.11$, $[1.03,1.20]$, $p=0.010$; $n=285$ subject-meetings). Excluding Area protection and management from the panel leaves the association stronger ($1.13$, $[1.03,1.24]$, $p=0.008$; $n=266$). A permutation null reassigns which subject's agreement series is paired with which subject's attention series across $200$ draws, holding each subject's own agreement and attention totals fixed; the observed Resolution coefficient exceeds the resulting null 95\% interval ($[-0.172,0.141]$ on the model scale, one-sided $p=0.020$). Subject-weighted counts of Measures show no association with later attention at any tested horizon (1, 3, 5, or 10 meetings). Subject-weighted counts of Decisions show a negative association that reaches $p<0.05$ only at the 3-meeting horizon and does not hold at the primary 5-meeting horizon or under the exclude-Area-protection subset. These associations do not establish that adopting an agreement causes subsequent attention. \url{scripts/analyze_output_to_subsequent_attention.py} reproduces the panel, the primary models, and both robustness checks.

\subsection{Measures cite earlier Measures most widely}\label{SI-sec:formal-inheritance-results}
The adopted body text of 245 of 277 Measures cites at least one earlier Measure, Decision, or Resolution (\cref{SI-tab:measure-formal-inheritance}). Prior Measures provide the broadest source set: 173 distinct earlier Measures appear in 474 links to 231 later Measures. The 261 Decision links come from only eight distinct Decisions, showing repeated use of a small administrative foundation. Nine distinct Resolutions appear in 81 links to 73 Measures.

\begin{table*}[!htbp]
\centering
\scriptsize
\caption{\textbf{Measures cite earlier Measures more widely than Decisions or Resolutions.} The inventory contains 277 Measures from regular ATCMs, 1995--2025. Links require an exact, year-qualified reference in a later Measure's adopted body text and a match to the official inventory. A Measure can cite several source types, so rows overlap. Lags are calendar years.}
\label{SI-tab:measure-formal-inheritance}
\begin{tabular}{@{}lrrrr@{}}
\toprule
Earlier agreement form & Distinct later Measures & Distinct earlier agreements & Citation links & Median lag \\
\midrule
Measure & 231 & 173 & 474 & 8 \\
Decision & 171 & 8 & 261 & 11 \\
Resolution & 73 & 9 & 81 & 18 \\
\bottomrule
\end{tabular}
\end{table*}

Seventy-eight of the 81 Resolution-to-Measure citation links have a target Measure in the official Area protection and management subject; two further links concern heritage work. Repeated Measures cite Resolutions that set expiry rules, management-plan formats, revision responsibilities, or heritage-assessment procedures. Resolution 4 (2007) on ship-based tourism and Measure 15 (2009) on passenger-vessel landings form the only cited pair outside those domains.

Exact citations miss intermediate steps that the final Measure does not repeat. Decision 3 (1998) moved further negotiation of a liability annex to ATCM Working Group I, and Resolution 5 (1999) requested practical and scientific input. Decision 3 (2001) advanced a draft, Measure 1 (2005) adopted Annex VI, and Decision 1 (2005) established annual review of progress towards its becoming effective. The 2005 Measure does not cite the 1999 Resolution. For this reason, \cref{SI-tab:measure-formal-inheritance} is a lower bound on documented continuity and not a complete reconstruction of how agreements develop.

Requiring ten subsequent meetings leaves 115 eligible Resolutions; Measures adopted within that horizon cite six. Their same-subject attention is not unusually high relative to uncited Resolutions sharing a subject. The linked Resolutions average the 24th percentile for the current and two preceding meetings, then the 33rd, 36th, 39th, 42nd, and 43rd percentiles after one, three, five, eight, and ten meetings. The sample is too small for a general transition model, but it does not support a simple rule in which Measures later cite the most heavily discussed Resolutions. A later Measure citation also does not establish that the Measure entered into force.

\subsection{The paper--agreement association changes with sample period}\label{SI-sec:attention-accumulation-results}
\begin{figure*}[!htbp]
\centering
\includegraphics[width=\textwidth]{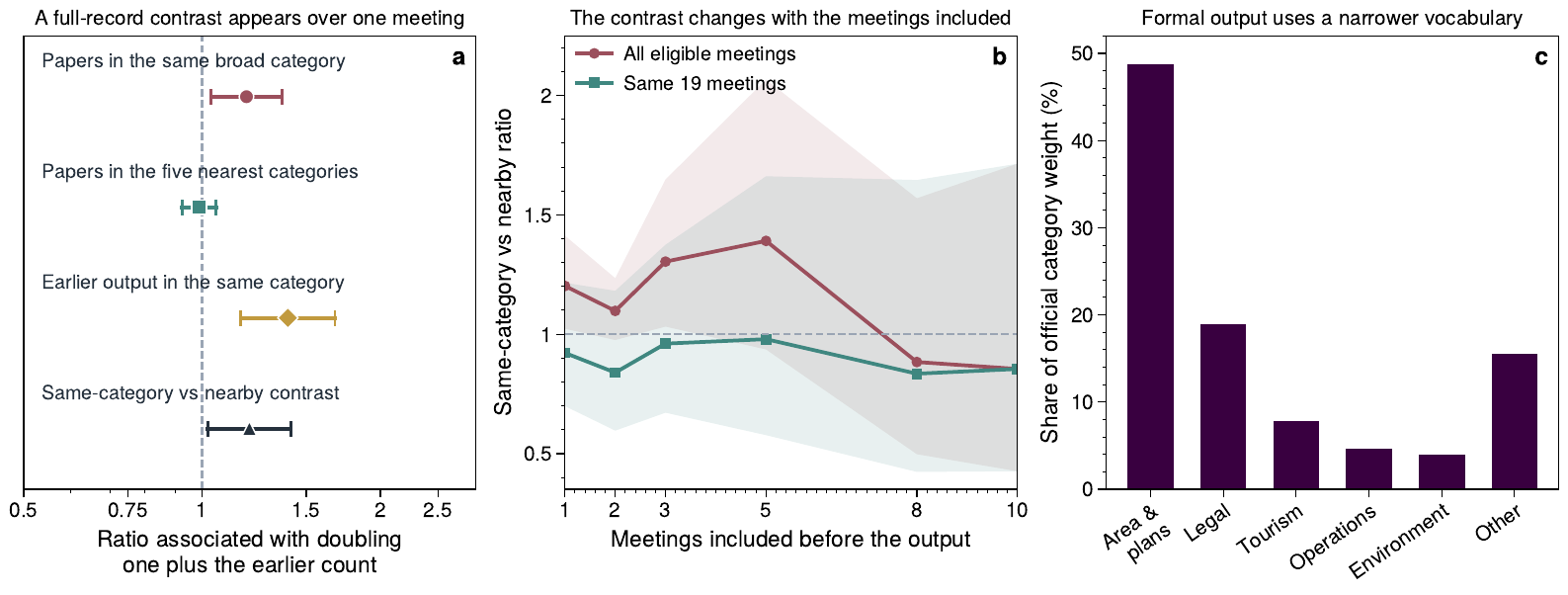}
\caption{\textbf{The paper--agreement association across forms of agreement changes with sample period.} (a) Agreement-rate ratios associated with doubling one plus an earlier count from the immediately preceding Antarctic Treaty Consultative Meeting (ATCM). The fourth estimate is the ratio of the same-subject and linked-subject rate ratios. Bars are subject- and meeting-clustered 95\% intervals. (b) The same ratio across windows of 1, 2, 3, 5, 8, and 10 meetings, using all eligible meetings or the common 19-meeting sample. (c) Fractional subject weight for 584 agreements across 15 subjects. This boundary test combines all three forms of agreement and differs from predictions fitted separately for each form.}
\label{SI-fig:pooled-output-boundary}
\end{figure*}

At the immediately preceding ATCM, doubling one plus papers in the same subject corresponds to an agreement-rate ratio of $1.189$ ($[1.036,1.364]$). The corresponding estimate for papers in the five most closely linked subjects is $0.989$ ($[0.926,1.055]$), and that for earlier agreements in the same subject is $1.396$ ($[1.162,1.677]$). The ratio between the same-subject and linked-subject rate ratios is $1.202$ ($[1.023,1.413]$, $p=0.025$; \cref{SI-fig:pooled-output-boundary}a). The linked-subject estimate is conditional on same-subject papers and should not be read as evidence that attention to linked subjects suppresses agreements.

Changing the window also changes which meetings can enter the model. With every eligible meeting included, the same-subject versus linked-subject contrast is clear at one meeting ($1.202$, $[1.023,1.413]$) and three meetings ($1.304$, $[1.032,1.648]$), but not at five meetings ($1.391$, $[0.936,2.067]$). Over the same 19 later meetings, the corresponding contrasts are $0.922$ ($[0.699,1.216]$), $0.961$ ($[0.671,1.375]$), and $0.979$ ($[0.577,1.662]$). None of the estimates over the same 19 meetings is clear (\cref{SI-fig:pooled-output-boundary}b). The full-record result therefore varies with sample period rather than identifying a fixed lag or cumulative conversion law.

Composition changes the precision of the result. After removing Area protection and management from the paper histories, coarser network, and agreement panel, the one-meeting contrast is $1.309$ ($[0.997,1.719]$); after also removing General matters, it is $1.230$ ($[0.853,1.774]$). The point estimates do not shrink, so these checks do not show that area management creates the association. They do show that the full-record contrast is no longer clear once we remove the dominant subject. Almost half of the subject weight of adopted agreements belongs to Area protection and management, much of it recurring site and management-plan work (\cref{SI-fig:pooled-output-boundary}c).

\subsection{Consultative Parties preserve the main patterns}\label{SI-sec:consultative-party-sensitivity}

We repeated the concern-network, new-specialization, and agreement-prediction analyses after restricting documentary attention to Consultative Parties. We used recognition dates from the official ATS Parties table and treated a state as eligible in its recognition year and every subsequent meeting year. Papers submitted only by Non-Consultative Parties, observers, invited expert organizations, or the Secretariat did not enter this analysis. For jointly submitted papers, only eligible Consultative Parties received participant-level credit, while each retained paper contributed one total unit of attention across its concerns. All other definitions matched the main analyses.

\begin{figure*}[!htbp]
\centering
\includegraphics[width=\textwidth]{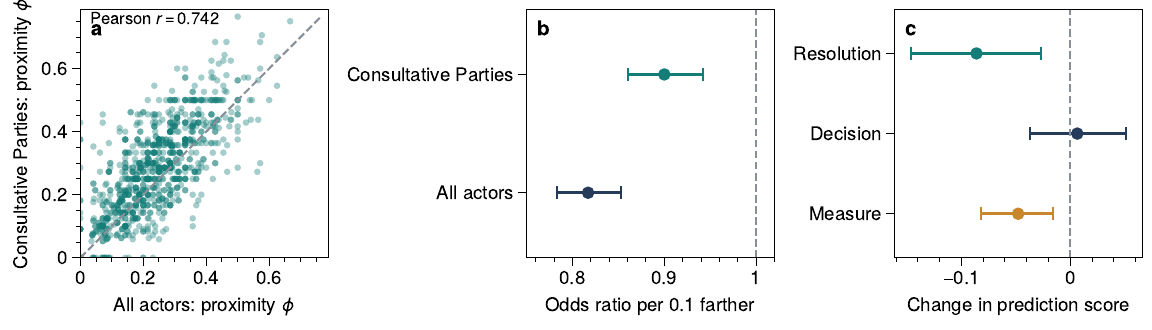}
\caption{\textbf{Consultative-Party restrictions preserve network connectivity, local portfolio development, and gains from documentary attention.} (a) Pairwise concern proximity in the full participant set and among Consultative Parties only. The dashed line marks equality. (b) Odds ratios for entering a specialization $0.1$ farther from the earlier portfolio, with model-based 95\% intervals. (c) Prediction-score change for the revised model with separate Working Paper and Information Paper counts and attention to linked subjects, relative to agreement history. These predictions are refitted for the restricted participant set. Negative values indicate improvement; bars show paired meeting-bootstrap 95\% intervals.}
\label{SI-fig:consultative-party-sensitivity}
\end{figure*}

The restricted corpus contains 5,192 papers attributed to 29 Consultative Parties. All 45 concerns remain connected, and pairwise proximities correlate with the all-participant reconstruction at Pearson $r=0.742$ and Spearman $\rho=0.730$ (\cref{SI-fig:consultative-party-sensitivity}a). New specializations remain local: each $0.1$ increase in distance corresponds to an odds ratio of $0.900$ (95\% interval $[0.860,0.942]$) across 741 participant-period comparisons (\cref{SI-fig:consultative-party-sensitivity}b). Rebuilding the separate-document-type prediction model gives a Resolution score change of $-0.086$ (95\% interval $[-0.146,-0.027]$), improving 12 of 19 meetings. The Measure change is $-0.048$ (95\% interval $[-0.082,-0.016]$) and the Decision change is $0.006$ (95\% interval $[-0.037,0.051]$), all relative to agreement history (\cref{SI-fig:consultative-party-sensitivity}c). Local portfolio development and the revised Resolution and Measure gains therefore persist under the restriction to formal decision-makers.

\subsection{Restricting attention to the original signatories}\label{SI-sec:original-signatory-sensitivity}

We also repeat the three analyses using a fixed set of original signatories: Argentina, Australia, Belgium, Chile, France, Japan, New Zealand, Norway, South Africa, the United Kingdom, the United States, and the Soviet Union, represented by the archive's Russian Federation participant label. This restriction retains 4,000 papers with at least one eligible participant. Only eligible participants receive participant-level credit, and each retained paper contributes one total unit across its concerns. We rebuild the concern networks and paper predictors for this participant set, keep all adopted agreements as prediction targets, and retain the settings used in the Consultative-Party-only analysis. This comparison asks whether the patterns persist without later entrants; it does not isolate the effect of gaining Consultative status.

All 45 concerns remain connected, with positive relations for 868 of 990 pairs. The detailed network nevertheless changes: pairwise scores correlate with the all-participant network at Pearson $r=0.472$ and Spearman $\rho=0.471$. New specializations remain local across 389 participant-period comparisons, with an odds ratio of $0.892$ per $0.1$ increase in distance (95\% interval $[0.847,0.939]$). Refitting the revised separate-document-type model gives a Resolution score change of $-0.086$ (95\% interval $[-0.143,-0.031]$), improving 13 of 19 meetings. The Measure change is $-0.043$ (95\% interval $[-0.082,-0.007]$) and the Decision change is $0.013$ (95\% interval $[-0.037,0.064]$). The additional contribution of linked subjects to Resolution prediction is $0.002$ (95\% interval $[-0.003,0.007]$). All comparisons retain all adopted agreements as targets. The detailed network changes even though local portfolio development and the Resolution and Measure gains persist. \url{scripts/analyze_original_signatory_sensitivity.py} reproduces the network and locality comparison; \url{scripts/analyze_document_type_prediction.py} produces the revised predictions (\cref{SI-fig:document-type-prediction}c).

\subsection{Cohort positions differ during the first 15 years}\label{SI-sec:secondary-structure-results}

In a retrospective comparison using the full-record network, participants joining in 1991--2010 ($n=17$) built opening portfolios around concerns whose specializing participants had smaller full-record portfolios than expected ($0.929\times$ the weighted comparison; 95\% participant-bootstrap interval $[0.889,0.966]$). Those concerns also have fewer closely linked concerns in that network on average ($0.883\times$ the comparison, $[0.798,0.974]$). The comparison weights concerns by their share of recorded specializations during each participant's first 15 years; both portfolio size and counts of closely linked concerns use full-record network scores. Earlier cohorts sit near parity on both measures. The full estimates and construction are in \cref{SI-sec:cohort-method,SI-tab:cohort-ratios}. This selected comparison excludes post-2010 entrants by design, conditions eligibility on a 15-year record, and does not reconstruct the information set available at entry. Equal spans of recorded activity do not give participants the same historical opportunities: the concerns facing the ATS differ across entry periods, which may contribute to the descriptive cohort differences.

\subsection{Alternative matched comparisons}\label{SI-sec:paired-null-robustness}
The paired test compares each new specialization with a size-matched redraw from concerns in which the participant did not specialize. We vary the full-record network or networks built from all earlier records, uniform or popularity-weighted redraws, and whether we draw candidates from all eventual labels, those observed by the current-window end, or those observed before the transition. Popularity weights equal the number of participants specialized in the concern in the previous window plus one. Each event uses 200 redraws.

Every construction places new specializations closer to the earlier portfolio than its matched alternatives more often than not. The primary comparison uses the network built from all earlier records, prior popularity, and only concerns already observed before the transition. It contains 968 participant-periods and gives $62.5\%$ closer (95\% participant-bootstrap interval $[58.9,66.5]\%$). The observed median displacement is $0.576$, against a weighted-alternative median of $0.594$ ($p=1.3\times10^{-15}$ in the unclustered Wilcoxon comparison). \url{scripts/compare_major_results_category_treatments.py} reproduces the comparison and participant bootstrap.

\begin{table*}[!htbp]
\centering
\scriptsize
\caption{\textbf{Local specialization persists across alternative definitions and timing choices.} Rows vary the comparison, outcome, category assignment, or time window. OR is the odds ratio for entering a specialization after a $0.1$ increase in distance from the earlier portfolio. Values below one indicate locality. The first row uses a participant-bootstrap interval. The complete-history bootstrap row reports the median and central 95\% percentile range of full-rebuild OR estimates; other OR rows use model-based 95\% intervals. In the combined robustness row, estimates report the first-paper outcome followed by one-category assignment. The final row reports duration under alternative lapse rules and has no OR.}
\label{SI-tab:locality_estimands}
\begin{tabular}{@{}p{0.30\textwidth}p{0.22\textwidth}p{0.40\textwidth}@{}}
\hline
\textbf{Comparison} & \textbf{Result} & \textbf{Question answered} \\
\hline
Popularity-weighted alternatives using only earlier information ($n=968$) & 62.5\% closer, participant-bootstrap interval $[58.9,66.5]\%$ & Is the new specialization closer to the earlier portfolio than alternatives weighted by their earlier popularity? \\
\hline
Within-participant model using only earlier information (30,957 participant-period--concern options) & OR $0.817$ $[0.783,0.853]$ per $0.1$ farther & Among options open to the same participant at the same time, are closely linked concerns more likely to cross the threshold? \\
\hline
Complete participant-history bootstrap (500 full rebuilds) & Median OR $0.869$; central 95\% range $[0.820,0.918]$ & Does locality persist when complete participant histories are resampled and the network and choice model are rebuilt? \\
\hline
First-paper outcome and one-category robustness & OR $0.874$ $[0.834,0.916]$ and $0.831$ $[0.794,0.870]$ & Does the direction remain after changing the outcome or forcing one inferred category per paper? \\
\hline
Disjoint five-meeting windows & OR $0.904$ $[0.855,0.957]$ & Does the adjacent-window result extend to non-overlapping blocks? \\
\hline
States-only choice sets (42 states; 821 participant-periods) & OR $0.827$ $[0.789,0.866]$ & Does locality persist when only states contribute choice sets, retaining the all-participant geometry? \\
\hline
Years a new specialization lasts & 5--42\% remain after ten years and fewer than 20\% after twenty years, depending on whether the rule allows a lapse of one to five years & How much does the length of apparent specialization depend on the rule for a temporary lapse? \\
\hline
\end{tabular}
\end{table*}

\FloatBarrier

\subsection{A boundary test: the 1991 Madrid Protocol}\label{SI-sec:madrid-shock}

To ask whether positions before the Madrid Protocol also predict changes beyond routine portfolio development, we examine the reorientation surrounding the 1991 Madrid Protocol. Parties adopted the Convention on the Regulation of Antarctic Mineral Resource Activities (CRAMRA) in Wellington in 1988 and abandoned it after Australia and France declined to sign it in 1989; the Protocol negotiations of 1989--91 replaced it \cite{Joyner1998}. The documentary record in 1986--90 therefore contains attention to the failed convention as well as the concerns later reorganized by the Protocol.

The design uses pre-1991 inputs where possible. We first restrict eligibility to concerns representing at least $1\%$ of the pre-1991 paper record, which excludes concerns created by the Protocol. Among the eligible concerns, the post-Protocol growth set contains the eight whose archive share grew most between 1986--90 and 1992--96. Using documents from 1961--90, we measure how close a participant's pre-1991 portfolio ($RPA \ge 1$) lies to the post-Protocol growth set by its highest proximity score. The outcome measures the share of a participant's papers in each 5-year window that falls in the post-Protocol growth set. We model it with window dummies interacted with this closeness measure, participant and window fixed effects, window-by-portfolio-size controls, and participant-clustered standard errors. The model covers 30 participants and 263 participant-window observations with at least three paper--concern records.

The pre-period coefficients do not reject parallel trends (joint $F(5,\cdot)=1.61$, $p=0.159$), although this low-powered diagnostic does not establish that the shock is exogenous. Every post-shock closeness-by-window coefficient is negative and small relative to its uncertainty (range $-0.83$ to $-0.19$ percentage points). A second-pass model asks whether relative closeness to the post-Protocol growth set predicts above-average within-window attention to any of six predefined Protocol concerns: Liability, Inspections, Environmental Monitoring and Reporting, Environmental Protection (General), Waste Management and Disposal, and Comprehensive Environmental Evaluations. The indicator equals one when at least one of these concerns has a paper count above the participant's mean count per concern addressed in that five-year window. It can equal one in successive windows and uses a different threshold from RPA. Its coefficient is also negative and uncertain ($-3.02$, SE $3.12$, $p=0.334$). The closeness measure correlates strongly with pre-1991 portfolio size ($r=0.87$), which the window-by-portfolio-size controls absorb.

The negative, imprecise estimates leave the association under an unusually large institutional reorientation unresolved. Pre-shock position did not clearly distinguish who responded, and the design cannot determine whether the collective response overrode a normally local process or arose through a different process altogether. We report this exploratory boundary test because its outcome limits the mechanism reading of the main text. \url{scripts/madrid_shock_exploration.py} reproduces the analysis.

\printbibliography

\end{document}